%% file: msV1.tex
\documentclass[twocolumn]{aastex631}
\received{June 10, 2022}
\revised{August 26, 2022}
\accepted{September 13, 2022}

\journalinfo{Accepted for publication in The Astronomical Journal}

\shorttitle{MAGIC}
\shortauthors{Zhao \& Zhu}
\graphicspath{{./}{figures/}}

\usepackage{CJK}

\begin{document}
\begin{CJK*}{UTF8}{gbsn}

\title{MAGIC: Microlensing Analysis Guided by Intelligent Computation}

\correspondingauthor{Wei Zhu (祝伟)}
\email{weizhu@mail.tsinghua.edu.cn}

\author[0000-0001-6675-1489]{Haimeng Zhao (赵海萌)}
\affiliation{Department of Astronomy, 
Tsinghua University, 
Beijing 100084, China}
\affiliation{Zhili College, 
Tsinghua University, 
Beijing 100084, China}

\author[0000-0003-4027-4711]{Wei Zhu (祝伟)}
\affiliation{Department of Astronomy, 
Tsinghua University, 
Beijing 100084, China}








\begin{abstract}
The modeling of binary microlensing light curves via the standard sampling-based method can be challenging, because of the time-consuming light curve computation and the pathological likelihood landscape in the high-dimensional parameter space. In this work, we present MAGIC, which is a machine learning framework to efficiently and accurately infer the microlensing parameters of binary events with realistic data quality. In MAGIC, binary microlensing parameters are divided into two groups and inferred separately with different neural networks. The key feature of MAGIC is the introduction of neural controlled differential equation, which provides the capability to handle light curves with irregular sampling and large data gaps. Based on simulated light curves, we show that MAGIC can achieve fractional uncertainties of a few percent on the binary mass ratio and separation. We also test MAGIC on a real microlensing event. MAGIC is able to locate the degenerate solutions even when large data gaps are introduced.
As irregular samplings are common in astronomical surveys, our method also has implications to other studies that involve time series.
\end{abstract}



\section{Introduction} \label{sec:intro}
\input{sec_intro}

\section{Binary Microlensing Parameters} \label{sec:param}
\input{sec_param}

\section{Inference of time-dependent parameters} \label{sec:unet}
\input{sec_unet}

\section{Inference of time-independent parameters} \label{sec:ncde}
\input{sec_ncde}

\section{Joint Pipeline \& Real Event Application} \label{sec:application}
\input{sec_application}


\section{Discussion} \label{sec:disc}
\input{sec_disc}

\begin{acknowledgments}
We thank Zerui Liu and Minghao Liu for discussions.
{We also thank the anonymous referee for comments and suggestions on a previous version of the manuscript.}
This work is supported by the National Science Foundation of China (grant Nos.\ 12173021 and 12133005) and the science research grant from the China Manned Space Project with No.\ CMS-CSST-2021-B12.
We also acknowledge the Tsinghua Astrophysics High-Performance Computing platform for providing computational and data storage resources. 
\end{acknowledgments}

%

\vspace{5mm}


\software{NumPy \citep{harris2020array}, SciPy \citep{2020SciPy-NMeth}, Pandas \citep{mckinney-proc-scipy-2010-pandas}, Matplotlib \citep{Hunter2007matplotlib}, PyTorch \citep{pytorch}, MulensModel \citep{mulensmodel}, corner \citep{corner}, Jupyter \citep{jupyter}, torchcde \citep{kidger_thesis}.}



\appendix

\section{Events with Large $\chi^2$} \label{app:param_vs_lc}
\input{app_large_chi2}


\bibliography{my_bib}{}
\bibliographystyle{aasjournal}



\end{CJK*}
\end{document}

%% file: sec_intro.tex
When a distant star (called the source) gets sufficiently aligned with a massive foreground object (called the lens), the gravitational field of the lens focuses the light out of the distant star, thus making the distant star appear brighter \citep{Einstein:1936, Paczynski:1986}. For a typical source star inside the Milky Way, one can observe the time evolution of their brightness (i.e., light curves) and infer the existence and properties of companion objects to the lens by monitoring the deviations in the light curve from the single lens scenario \citep[e.g.,][]{Mao:1991, Gould:1992}. This so-called gravitational microlensing technique has been frequently used to detect exoplanets and stellar binaries and are complementary to other techniques (see reviews by \citealt{gaudi2012microlensing} and \citealt{Zhu:2021}).

For a microlensing event with multiple lenses, the interpretation of the light curve can be challenging. First of all, the computation of the multiple-lens microlensing light curve can be time-consuming due to the finite-source effect \citep[e.g.,][]{Dong:2006, Bozza:2010}. This is especially true when the microlens system consists of three or more objects \citep[e.g.,][]{Gaudi:2008, Kuang:2021}. Additionally, the likelihood landscape of the high-dimensional parameter space can be so pathological that traditional sampling-based methods may have a hard time searching for the correct solution (or solutions). This remains to be true even when the brute force search on a fine grid that is defined by a subset of model parameters is conducted. As a result, the current analysis of multiple-lens microlensing events is still case-by-case, with each event requiring hundreds of (or more) CPU hours as well as (usually) the supervision of domain experts \citep[e.g.,][]{Dominik:1999, GouldHan:2000, An:2005, Song:2014, Zang:2022, Zhang:2022}.

Several methods have been proposed to automate the parameter estimation problem in microlensing events with two lenses \citep{verm2003rapid, Khakpash:2019, zhang2021real, Kennedy:2021}. These binary-lens events are the overwhelming majority of all anomalous events in current microlensing surveys, as events with more lenses are expected to be rare \citep{Zhu:2014}. Of the proposed methods, machine learning is a promising approach that may efficiently and accurately infer parameters out of binary microlensing light curves \citep{verm2003rapid, zhang2021real}, in addition to its application to the identification (or classification) of microlensing events \citep{wyrzykowski2015ogle, GODINES2019100298, mroz2020ident}. However, the existing machine learning methods for parameter inferences are not readily applicable to microlensing data from the ongoing surveys. In particular, \citet{zhang2021real} trained and evaluated their deep learning network on simulated light curves from the \emph{Roman} microlensing survey \citep{Penny:2019}. While their model can successfully predict the posterior distributions of microlensing model parameters and identify degenerate solutions \citep{Zhang:2022}, their simulated light curves contain $\sim10^4$ time stamps with equal steps and high photometric precision. Ongoing microlensing surveys from the ground, such as the Korea Microlensing Telescope Network \citep[KMTNet,][]{Kim:2016}, see light curves with lower signal-to-noise (S/N) ratios and irregular samplings (including data gaps due to seasonal, diurnal, and instrumental issues). These realistic features cannot be handled in the framework of \citet{zhang2021real}.

Irregular time series are common in astronomical observations from the ground (and sometimes from space as well). These irregular time steps pose challenges to the family of recurrent neural networks \citep[RNNs,][]{goodfellow2016deep}, which is widely considered the standard machine learning method to deal with time series data. Variants of RNNs have been invented to handle irregular time series \citep{che2018recurrent}. For example, \citet{Charnock:2017} imputed/interpolated on the originally unevenly sampled light curves and obtained evenly sampled ones, whereas \citet{Naul:2018} added the time steps between consecutive observations as additional information into the standard RNN. Neither of these approaches would work well in the present situation. {As the source enters or exits} the caustic curve of the binary lens, microlensing light curves may change behaviours within brief time intervals. If these intervals fall into the data gap, it will be difficult for the standard interpolation methods to make up the missing features without introducing false signals \citep{rubanova2019latent}.

\begin{figure*}
    \centering
    \includegraphics[width=0.9\linewidth]{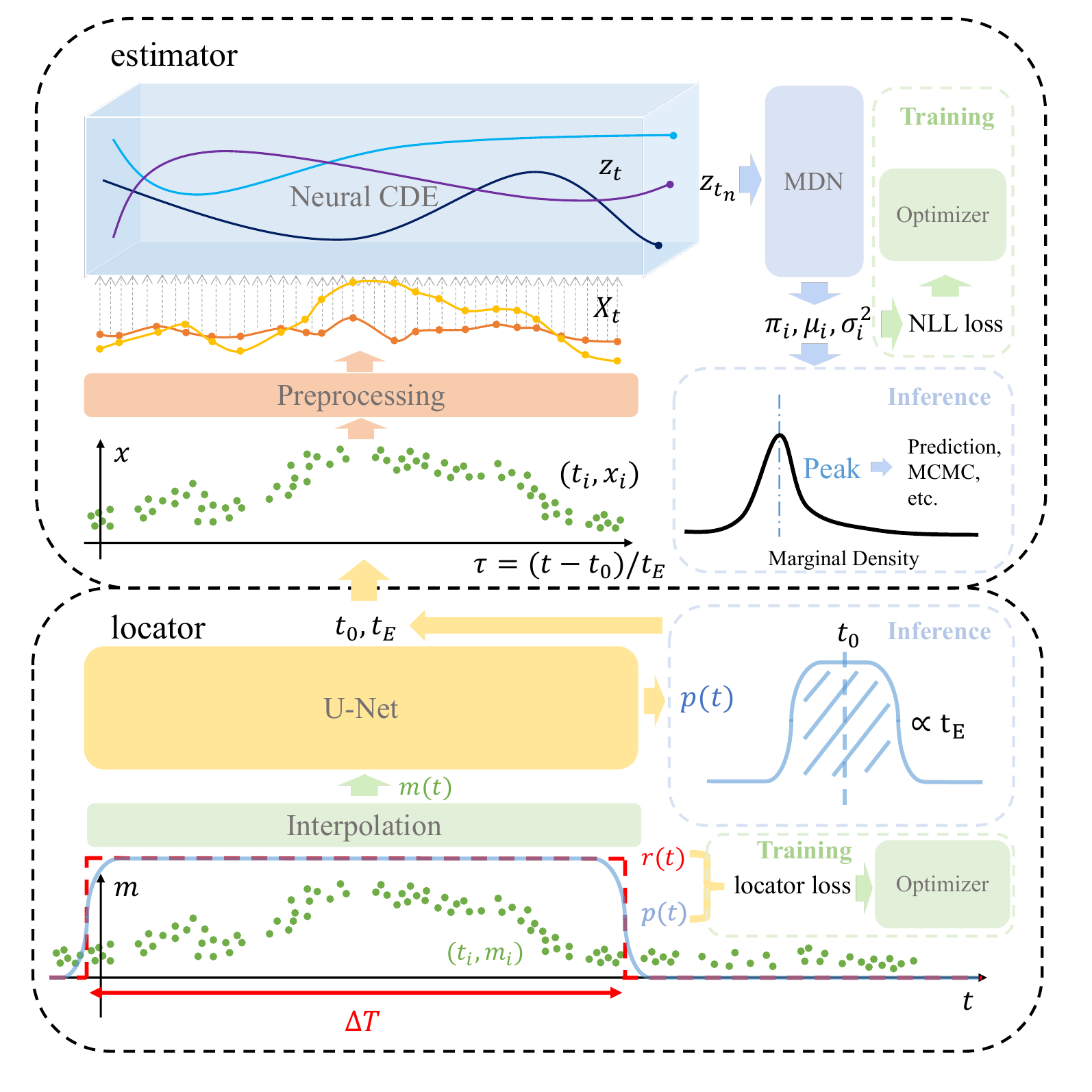}
    \caption{The schematic view of MAGIC. The binary microlensing parameters are divided into two groups. We use \texttt{locator}, which is a U-Net architecture, to infer the time-dependent parameters, namely the peak time $t_0$ and the event timescale $t_{\rm E}$. The output is used to transform the light curve into a function of $\tau \equiv (t-t_0)/t_{\rm E}$, and \texttt{estimator}, which combines neural controlled differential equation (neural CDE) and mixture density network (MDN), is used to infer the other microlensing parameters. We refer to Sections~\ref{sec:unet}, \ref{sec:ncde}, and \ref{sec:application} for the detailed description of our method. } 
    \label{fig:model}
\end{figure*}

In this paper, we present Microlensing Analysis Guided by Intelligent Computation, or MAGIC for short. MAGIC is a machine learning framework that can efficiently and accurately perform parameter inference in binary microlensing events with realistic sampling conditions. A schematic overview of the proposed framework is shown in Figure~\ref{fig:model}. The binary microlensing parameters are explained in Section~\ref{sec:param}. We divide these parameters into two groups and develop different machine learning methods in Sections~\ref{sec:unet} and \ref{sec:ncde}. Section~\ref{sec:application} describes the joint pipeline and its application to a real microlensing event. Finally, we discuss our results in Section~\ref{sec:disc}.

%% file: sec_param.tex
We describe the parameters that are required to model the binary microlensing events. These parameters are grouped and inferred separately.

The standard binary microlensing light curve can be defined by the following parameters {(see \citealt{gaudi2012microlensing} and references therein)}
\begin{equation} \label{eqn:all-parameters}
    (t_0, t_{\rm E}, u_0, \rho, q, s, \alpha, m_{\rm base}, f_{\rm S}) .
\end{equation}
Here $t_0$ is the time of the closest approach between the source and lens under the rectilinear trajectory, $u_0$ is the separation between the source and the lens at $t_0$, $\rho$ is the size of the source star, and $t_{\rm E}$ is the event timescale
\begin{equation}
    t_{\rm E} \equiv \frac{\theta_{\rm E}}{\mu_{\rm rel}} ,
\end{equation}
where $\mu_{\rm rel}$ is the relative proper motion between the lens and the source. The Einstein ring radius $\theta_{\rm E}$ is given by
\begin{equation}
    \theta_{\rm E} \equiv \sqrt{\kappa M_{\rm L} \pi_{\rm rel}} ,
\end{equation}
where $\kappa \equiv 4G/{c^2 {\rm au}} \approx 8.14\,{\rm mas}/M_\odot$, $\pi_{\rm rel} \equiv {\rm au} (D_{\rm L}^{-1}-D_{\rm S}^{-1})$, with $M_{\rm L}$ and $D_{\rm L}$ the mass of and distance to the lens object/system, respectively, and $D_{\rm S}$ the distance to the source. Both $u_0$ and $\rho$ are normalized to the Einstein radius $\theta_{\rm E}$.

Three of the remaining parameters, $(q, s, \alpha)$, characterize the binary component. Here $q$ is the mass ratio between the secondary and the primary components, $s$ is the projected separation between the two binary components normalized to $\theta_{\rm E}$, and $\alpha$ measures the angle from the binary axis to the direction of the lens--source relative motion. We follow the convention of \citet{Skowron:2011} in the definition of the microlensing parameters except for $\alpha$, for which our definition is offset by $180^\circ$.
\footnote{This is the same in \texttt{MulensModel} \citep{mulensmodel}.}

The baseline magnitude of the microlensing light curve is given by $m_{\rm base}$, and the fraction of the total baseline flux coming from the microlensing source is given by $f_{\rm S}$. That is,
\begin{equation}
    m(t) = m_{\rm base} - 2.5 \lg[f_{\rm S} \cdot A(u_x, u_y) + 1-f_{\rm S}] ,
\end{equation}
where $A(u_x, u_y)$ is the light curve in microlensing magnifications. 
{The quantity, $1-f_{\rm S}$, represents the blending fraction, namely the fraction of total flux coming from stars/objects other than the source.}
The parameters $u_x$ and $u_y$ measure the displacement from the center of mass of the binary (i.e., the chosen origin) to the source at time $t$. In the chosen reference frame,
\begin{equation} \label{eqn:tau}
    \left( \begin{array}{c}
         u_x  \\
         u_y 
    \end{array} \right) = \left( \begin{array}{lr}
        \cos\alpha & -\sin\alpha \\
        \sin\alpha & \cos\alpha
    \end{array} \right) \left( \begin{array}{c}
        \tau  \\
        u_0
    \end{array} \right) ;\quad
    \tau \equiv \frac{t-t_0}{t_{\rm E}} .
\end{equation}

In principle, since microlensing observations are usually taken by telescopes at different sites or in different filters, each event may have multiple sets of $m_{\rm base}$ and $f_{\rm S}$ parameters. In practice, as long as the different datasets cover a broad range of microlensing magnifications, they can all be aligned to a single dataset to relatively high precision without reference to specific microlensing models \citep[e.g.,][]{Yoo:2004, Gould:2010a}. We therefore only include one set of $m_{\rm base}$ and $f_{\rm S}$ in the present work.

The binary microlensing parameters, given by Equation~(\ref{eqn:all-parameters}), are treated/inferred differently in the present work. First of all, it is reasonable to assume that the baseline magnitude $m_{\rm base}$ can be immediately read out of the ground-based light curve, which usually monitors (or can monitor) the source star for a time much longer than the event timescale. We therefore do not include $m_{\rm base}$ in our method. The scaled source size, $\rho$, is constrained only when the source is very close to or crosses the caustic curve of the binary lens. For the chosen (effective) observational cadence of this work, the finite-source effect is not expected to be widely detected. We therefore fix $\rho=10^{-3}$ and do not infer it. As will be shown later, our inference of the other parameters is not affected by whether or not the finite-source effect is present.

Of the remaining parameters, $t_0$ and $t_{\rm E}$ define the position and extent of the light curve in time, whereas the other parameters, $(u_0, q, s, \alpha, f_{\rm S})$, concern primarily the shape of the binary microlensing light curve. We show in Sections~\ref{sec:unet} and \ref{sec:ncde} that these two groups of parameters can be inferred separately. It is also worth noting that, once the other parameters are chosen, the source flux parameter, $f_{\rm S}$, can be derived analytically via the maximum likelihood method \citep{Gould:1995}. In Section~\ref{sec:ncde}, we make use of this property in the further optimization of the microlensing parameters.

Throughout this work, the \texttt{VBBinaryLensing} algorithm \citep[VBBL,][]{Bozza:2010, Bozza:2018} inside the \texttt{MulensModel} \citep{mulensmodel} package is used to generate binary microlensing light curves. We describe the generation of the training data set in the two different networks separately.

%% file: sec_unet.tex
\subsection{Method}

Together with $u_0$, parameters $t_0$ and $t_{\rm E}$ are the so-called Paczynski parameters \citep{Paczynski:1986}. Unlike the parameters that define the binary properties, these Paczynski parameters are constrained by the global properties of the light curve. In many low-$q$ events, especially planetary events, one can accurately constrain the Paczynski parameters by masking out the relatively brief anomalous features \citep[e.g.,][]{Gould:1992}. Of all three Paczynski parameters, the time-dependent ones, namely $t_0$ and $t_{\rm E}$, enter the microlensing phenomenon via the parameter $\tau$ (see Equation~\ref{eqn:tau}). For these reasons, we develop a stand-alone machine learning method to infer the time-dependent parameters, which is dubbed \texttt{locator}.

The basis of \texttt{locator} is the U-Net architecture. U-Net is a fully convolutional neural network that is originally designed to perform semantic segmentation of images \citep{ronneberger2015unet}. U-Net takes the original image as the input, and outputs a mask of the same shape. The mask can be interpreted as a pixel-wise classification result, representing, for example, the probability of each pixel belonging to some specific segment. U-Net has been ubiquitously used in biomedical image processing \citep{Siddique2021unet_review} since it was proposed in 2015. It has also seen a growing number of applications in astronomy \citep[e.g.,][]{VanOort:2019, Makinen:2021}. 

{The architecture of our \texttt{locator} closely follows that of \citep{ronneberger2015unet}. See their Figure~1 for an illustration. In details, \texttt{locator} differs in three aspects. First, all convolutional layers are one dimensional, because \texttt{locator} deals with time series data rather than image data. Second, the pure convolutional blocks are replaced by residual blocks \citep{he2016deep} with Parametric Rectified Linear Unit (PReLU, \citealt{he2015prelu}) activation functions to ease gradient back-propagation. Third, we set the channel and kernel sizes to be 128 and 7, respectively, to better match the shape of the input data. In practice, the performance of \texttt{locator} is not sensitive to the values of these hyper-parameters.}

We design \texttt{locator} to identify the light curve segment that falls within a predefined time window $\Delta T \equiv [t_0-k t_{\rm E}, t_0+k t_{\rm E}]$. Here $k$ is a parameter that can be tuned to optimize the output performance, and its choice will be discussed in Section~\ref{sec:unet_perform}. For each light curve, which is irregularly sampled (see Section~\ref{sec:unet_data}), we first interpolate it with natural cubic splines and extract the values at $n_t=4000$
\footnote{{This value of $n_t$ is chosen to be larger than the number of time stamps (2000, see Section \ref{sec:unet_data}). The resulting time resolution, $270/4000\approx 0.07$ days, is much smaller than the RMSE (2.2 days, see Section~\ref{sec:unet_perform}) of $t_0$.}}
equally stepped time stamps $\{t_{j}\}_{j=1}^{n_t}$.
\footnote{This crude approach of directly interpolating the irregular time series is justified given that the temporal parameters depend mostly on the macroscopic behaviour of the light curve, rather than the fine structure within a small time interval. This assumption is no longer valid for the parameters discussed in Section \ref{sec:ncde}.}
These time stamps are assigned labels depending on whether or not they are inside $\Delta T$,
\begin{equation}
    r_j = \left\{ \begin{array}{ll}
        1, & t_j \in \Delta T \\
        0, & {\rm otherwise} \\
    \end{array} \right. .
\end{equation}
At run-time, \texttt{locator} takes the time series and attempts to produce a mask $\{p_j\}_{j=1}^{n_t}$ that best approximates the ground truth labels.

We train \texttt{locator} on simulated light curves that match the key properties of real microlensing light curves (see Section~\ref{sec:unet_data}). The loss function that \texttt{locator} minimizes is given by
\begin{equation} \label{eqn:unet_loss}
    \mathcal{L}_{\text{locator}} 
    = \mathcal{L}_{\text{BCE}} + \mathcal{L}_{\text{Dice}} + \mathcal{L}_{\text{reg}}.
\end{equation}
Here $\mathcal{L}_{\rm BCE}$ is the binary cross entropy (BCE) loss function
\begin{equation}
    \mathcal{L}_{\text{BCE}} \equiv \mathop{\mathbb{E}}
    [r \ln p + (1 - r)\ln (1 - p)],
\end{equation}
{which is the standard loss function for binary classification \citep{goodfellow2016deep}.}
The expectation runs over all time stamps of all light curves in the training set. The second term in Equation~(\ref{eqn:unet_loss}) is the Dice loss function \citep{sudre2017dice}, which is introduced to assuage the class imbalance problem in events with small $t_{\rm E}$:
\begin{equation}
    \mathcal{L}_{\text{Dice}} \equiv \mathop{\mathbb{E}}
    \left[ 1 - \frac{\sum_{j} r_j p_j}{\sum_{j} (r_j + p_j)}
    \right] .
\end{equation}
The third term in Equation~(\ref{eqn:unet_loss}) is the regularization term to prevent the mask from wiggling up and down for more than twice
\begin{equation}
\mathcal{L}_{\text{reg}} \equiv \mathbb{E}
    \left(\sum_{j=1}^{n_t-1}|p_{j+1}-p_{j}| - 2\right)^2.
\end{equation}
The expectations in the last two terms are performed on all light curves in the training set.

The output mask from \texttt{locator} contains information about the event peak time, $t_0$, and the timescale, $t_{\rm E}$. Specifically, the centroid of the mask corresponds to $t_0$ and the total enclosed area is directly related to $t_{\rm E}$. In the continuum form, they are given by
\begin{equation} \label{eqn:t0te}
    t_0 = \frac{\int p(t) t dt}{\int p(t) dt}, \quad
    t_{\rm E} = \frac{1}{2k} \int p(t) dt .
\end{equation}
These integrals are evaluated via the trapezoidal rule in the discrete time series.

\subsection{Data generation} \label{sec:unet_data}

To simulate light curves for the training of \texttt{locator}, we randomly sample 2000 time stamps out of a total duration of $270\,$days. This corresponds to an averaged cadence of $\sim3\,$hr,
{with the lower and upper 10th percentiles  corresponding to 20\,min and 7.2\,hr.}
For each light curve, we randomly draw $t_{\rm E}$ from a truncated log-normal distribution with mean of $10^{1.15}\,$days, standard deviation of $10^{0.45}\,$days, and boundaries at 5 and $100\,$ days. This log-normal distribution matches approximately the observed distribution of $t_{\rm E}$ \citep[e.g.,][]{Mroz:2017}, and the truncations are introduced to eliminate relatively short-timescale and long-timescale events, both of which may be subject to additional issues (e.g., parallax effect for long events) and are not the focus of the present work. As for $t_0$, even though it can be any value during (and possibly outside) the sampled time interval, it is reasonable to believe that a very rough by-eye estimate can limit $t_0$ to within $\sim \pm t_{\rm E}$ of the truth. We therefore randomly draw $t_0$ out of a uniform distribution between $-t_{\rm E}$ and $t_{\rm E}$. To summarize,
\begin{eqnarray*}
    &t_{\rm E}/{\rm day} &\sim \text{TruncLogNorm}(5, 100, \mu=10^{1.15}, \sigma=10^{0.45}),\\
    &t_0 &\sim \text{Uniform}(-t_E, t_E).
\end{eqnarray*}

Although the goal of \texttt{locator} is to infer $t_0$ and $t_{\rm E}$ and not the other parameters, the training data set should be representative of the target---the binary light curves. We therefore sample the remaining parameters in the following way
\begin{eqnarray*}
&u_0 &\sim \text{Uniform}(0,1), \\
&q &\sim \text{LogUniform}(10^{-3}, 10^0), \\
&s &\sim \text{LogUniform}(0.3, 3), \\
&\alpha &\sim \text{Uniform}(0, 360), \\
&f_{\rm S} &\sim \text{LogUniform}(10^{-1}, 10^0),
\end{eqnarray*}
and $\rho$ is fixed at $10^{-3}$, given that the finite-source effect is not expected to be detected for the adopted (effective) cadence. Note that the mass ratio $q$ is limited to $>10^{-3}$ simply because we want to focus on relatively massive binaries. With increased sampling cadence \texttt{locator} 
{can in principle be extended down to lower $q$ values. We defer a detailed exploration to future works.} 

With the sampled binary parameters and time stamps, we calculate the microlensing light curve using the \texttt{VBBL} algorithm \citep{Bozza:2010, Bozza:2018}. To introduce noises to the simulated light curves, we adopt a simplified noise model. Gaussian noises with signal-to-noise ratios (S/N) of 33 in flux are assumed for all data points, and the flux values are converted to magnitudes for a baseline object with $m_{\rm base}=18$. 

Simulated light curves without significant binary features, defined as $\chi^2_{\rm pspl}$/d.o.f.\ $<1.25$,
\footnote{Here ``pspl'' represents for point-source point-lens.}
are rejected. In the end, two data sets each containing $10^5$ light curves are generated, one for training and the other for validation and test.

\begin{figure}
    \centering
    \includegraphics[width=\linewidth]{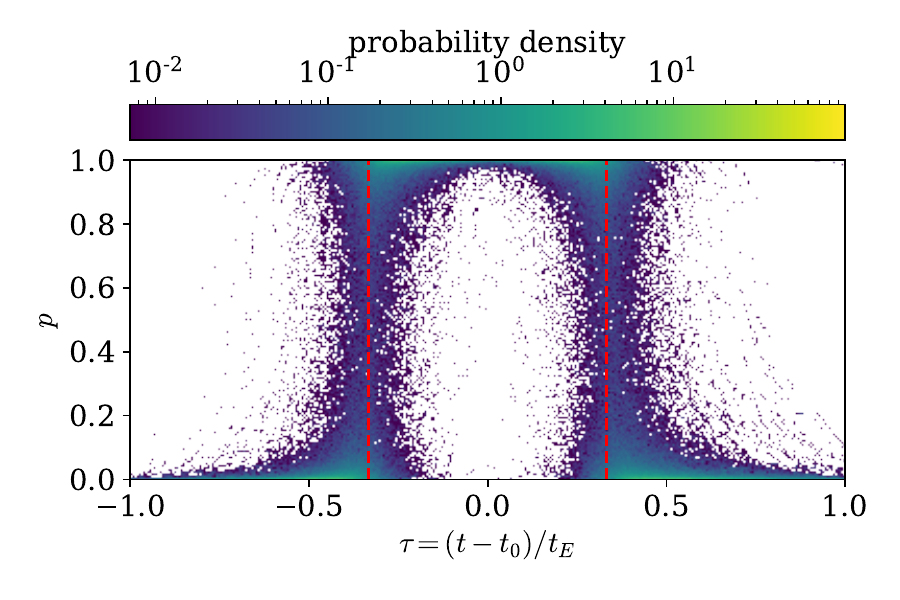}
    \caption{An illustration of the output mask of \texttt{locator}. The two-dimensional histogram shows the distribution of $p$ values inside the chosen $\tau$ window for 16,384 light curves with $s<1$. The red dashed lines indicate $\tau= \pm k = \pm 1/3$, which are the boundaries of the input mask in this realization. } 
    \label{fig:p_dist}
\end{figure}

\begin{figure}
    \centering
    \includegraphics[width=\linewidth]{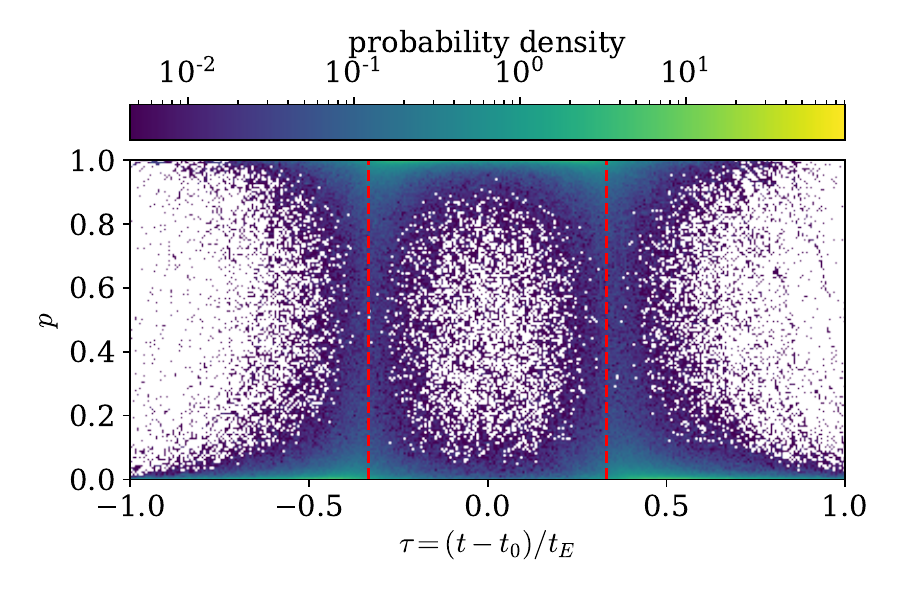}
    \caption{{Similar to Figure \ref{fig:p_dist}, but for events with $s>1$.}}
    \label{fig:p_dist_larges}
\end{figure}

\begin{figure*}
    \centering
    \includegraphics[width=0.8\textwidth]{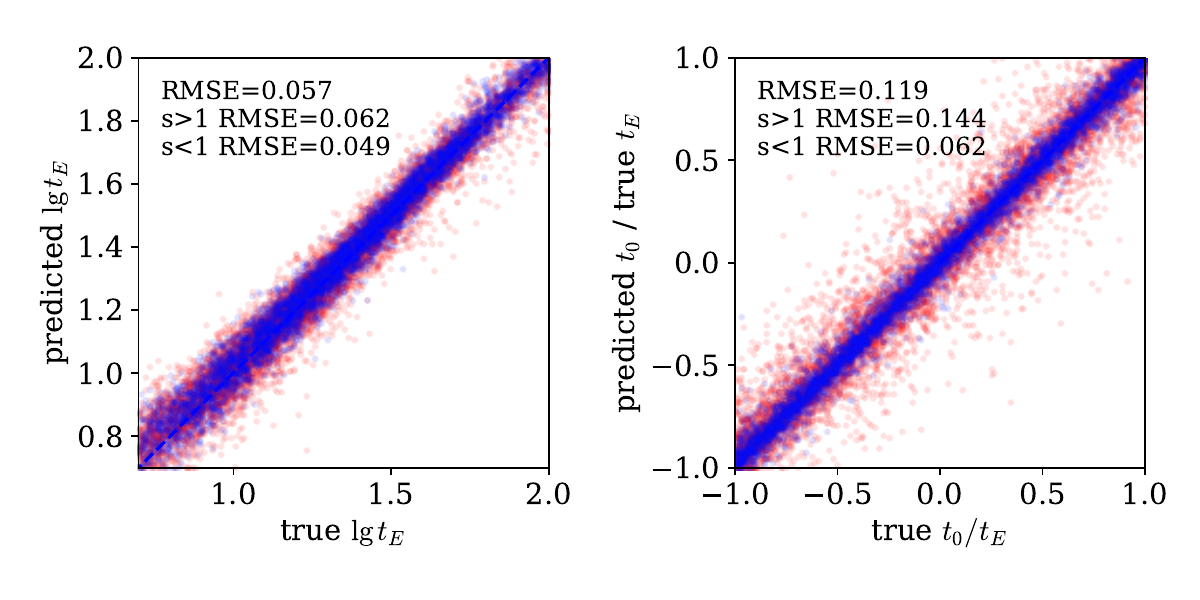}
    \caption{Comparisons between input and predicted values of $t_{\rm E}$ (left) and $t_0/t_{\rm E}$ (right). Values from events with $s>1$ and $s<1$ are separated, with the former shown in red and the latter in blue. The rooted mean square errors (RMSEs) of the parameters are computed and indicated at the upper left corners. In this realization, $k=1/3$ is used.}
    \label{fig:t0te}
\end{figure*}

\begin{figure*}
    \centering
    \includegraphics[width=\linewidth]{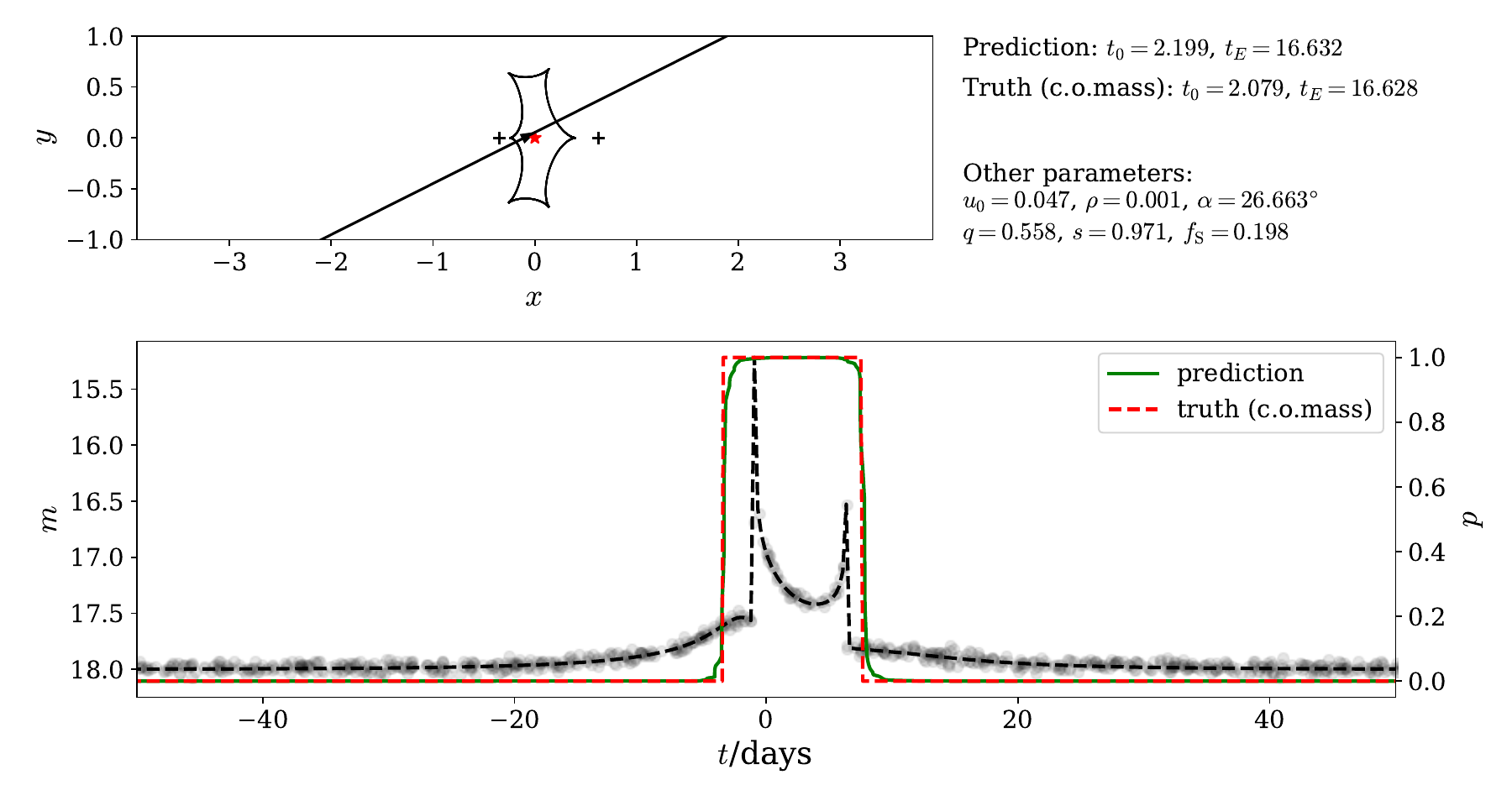}
    \caption{An example event in which \texttt{locator} successfully reproduces the input window and thus recovers $t_0$ and $t_{\rm E}$. The upper panel shows the microlensing geometry, with the red asterisk marking the center of mass (c.o.mass) of the binary lens. The lower panel shows the simulated light curve and the input and the output masks. The parameters used to reproduce this event are given on the upper right corner.
    \label{fig:lc_caustic}}
\end{figure*}

\subsection{Performance of \texttt{locator}} \label{sec:unet_perform}

For our default run, we set $k=1/3$ so that $\Delta T$ with the longest timescale ($100\,$days) can still be fully covered in the simulation interval $[-135, 135]$. We train \texttt{locator} with the ADAM optimizer \citep{kingma2014adam}. The learning rate is set to $4\times 10^{-3}$ initially and drops by ten percent after each epoch. Training samples are divided into mini-batches of size 128. After 13 epochs ($\sim$2\,hours) of training on one NVIDIA Tesla V100 GPU, the average loss drops to 0.16 on the validation set of size 1024. The optimized model is then used to predict the mask and estimate $t_0$ and $t_{\rm E}$ according to Equation~(\ref{eqn:t0te}). 

We evaluate \texttt{locator} on 16,384 light curves from the test set. The output masks are illustrated in Figure~\ref{fig:p_dist} {and Figure~\ref{fig:p_dist_larges}}, and the comparisons between the predicted and ground truth values of $\lg t_{\rm E}$ and $t_0/t_{\rm E}$ are shown in Figure \ref{fig:t0te}. 
The accuracy of the prediction is measured by the root-mean-squared error (RMSE) between the ground truth and the prediction. They are indicated in the upper left corner respectively.
We note that \texttt{locator} has very different performances on close ($s<1$) binaries and wide ($s>1$) binaries. This is especially true for the $t_0/t_{\rm E}$ parameter. We take the results of close binaries as a more reliable evaluation of the overall performance and defer to the end of this section a more detailed explanation of this phenomenon.

In the default run, \texttt{locator} achieves an accuracy in $\lg t_{\rm E}$ with RMSE$=0.049$, which translates to a fractional uncertainty of $11.3\%$ on $t_{\rm E}$. The accuracy on the ratio $t_0/t_{\rm E}$ is comparable at RMSE($s<1$)=0.062, which corresponds to an uncertainty of $1.9\,$days on $t_0$ for a typical event with $t_{\rm E}=30\,$days. An example event is shown in Figure~\ref{fig:lc_caustic}, in which the light curve substantially deviates from the \citet{Paczynski:1986} model due to the caustic crossing feature and yet \texttt{locator} successfully recovers $t_0$ and $t_{\rm E}$ at high accuracy.

\begin{figure*}
    \centering
    \includegraphics[width=0.8\linewidth]{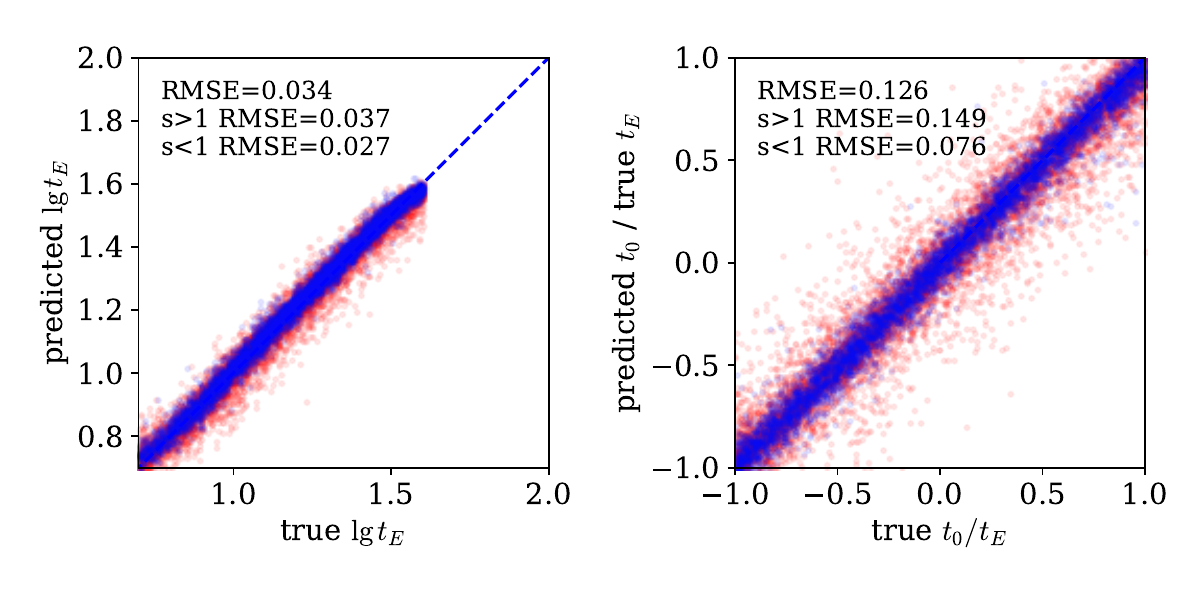}
    \caption{Similar to Figure~\ref{fig:t0te} except that only events with $t_{\rm E}<40\,$days and $k=2$ are used.}
    \label{fig:t0te_small}
\end{figure*}

\begin{figure*}
    \centering
    \includegraphics[width=\linewidth]{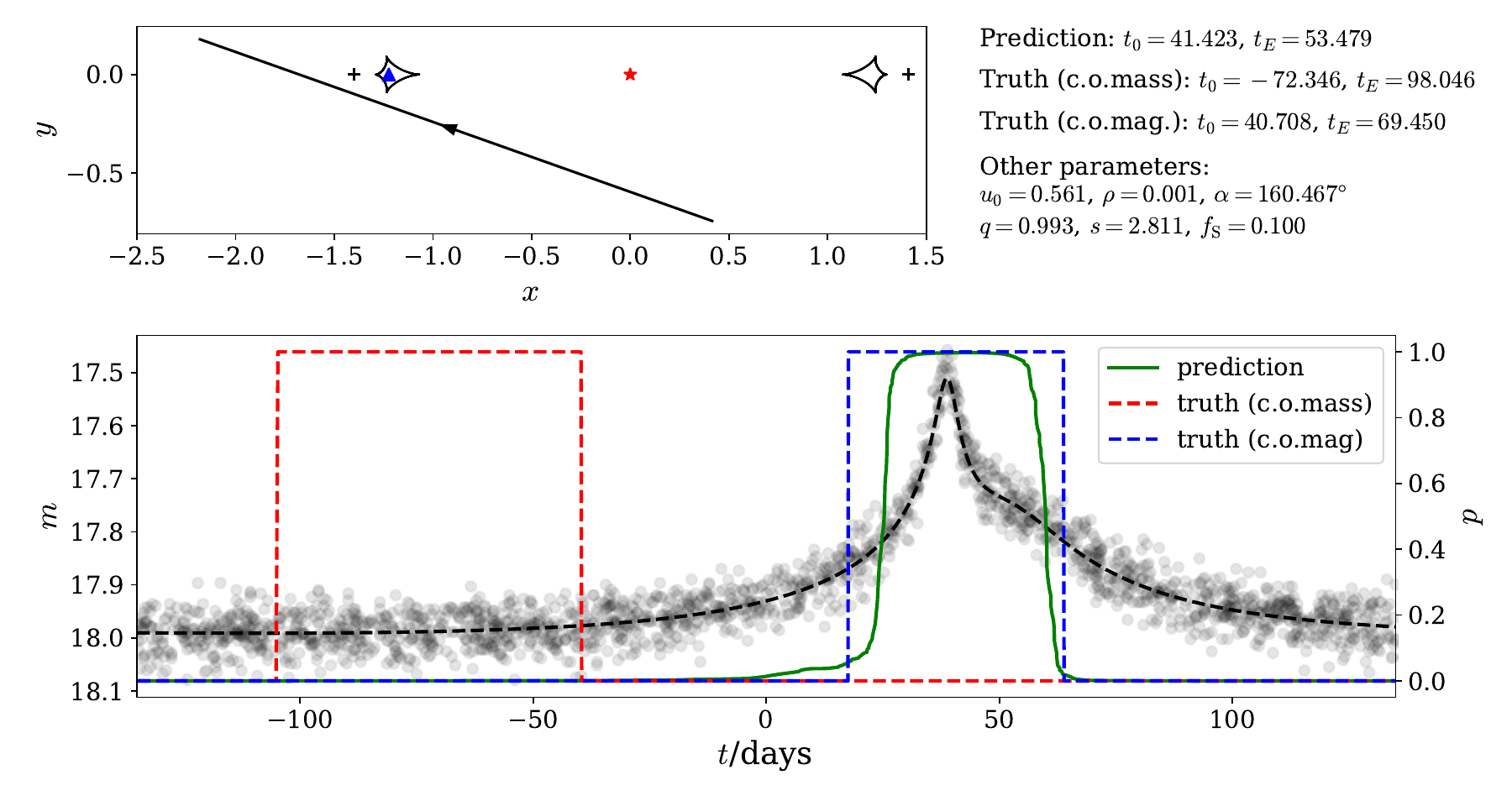}
    \caption{An example event with $s>1$, in which \texttt{locator} ``fails'' to reproduce the input window.
    The input window (red dashed curve) is defined on the center of mass (red asterisk in the upper panel) of this wide binary system.
    The prediction of \texttt{locator} (shown in green) better matches the light curve window (blue dashed curve) defined on the center of magnification (blue {triangle} in the upper panel) of the primary star (black plus on the left in the upper panel, with the one on the right being the secondary). This latter window better describes the time evolution of this wide binary event. 
    \label{fig:lc_s_wide}}
\end{figure*}

\begin{figure}
    \centering
    \includegraphics[width=0.9\linewidth]{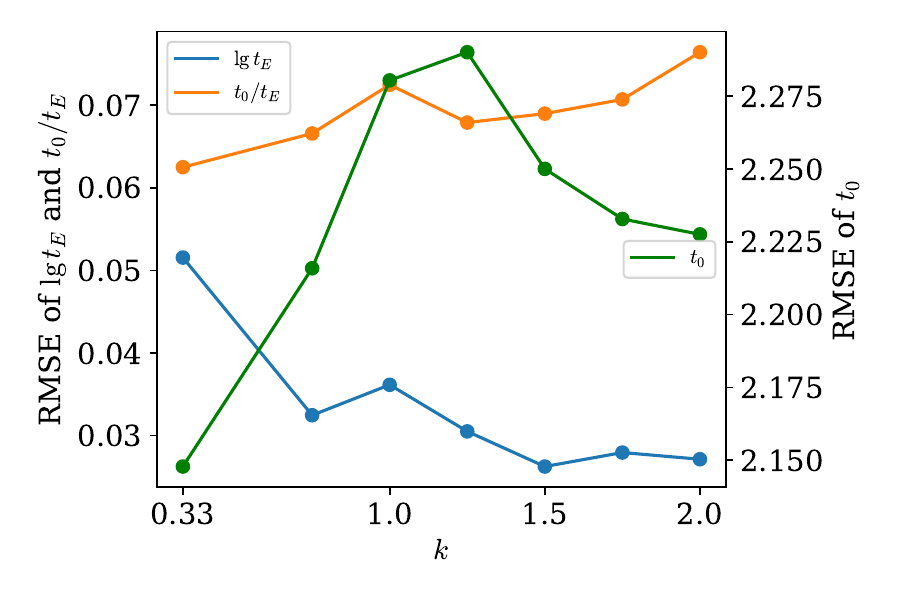}
    \caption{The performance of \texttt{locator}, evaluated on RMSEs of $\lg{t_{\rm E}}$, $t_0/t_{\rm E}$ and $t_0$, as a function of $k$ values. }
    \label{fig:rmse_k}
\end{figure}

\begin{figure}
    \centering
    \includegraphics[width=0.7\linewidth]{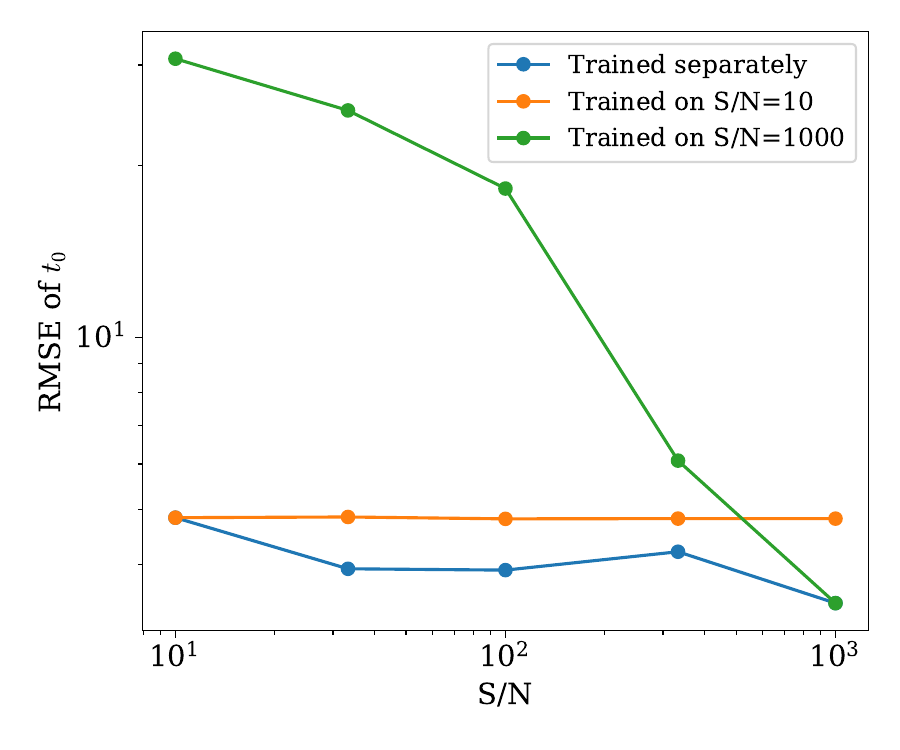}
    \caption{{The performance of \texttt{locator}, evaluated on RMSE of $t_0$, as functions of the photometric uncertainty (S/N) of individual data point. Three different training strategies are used. In blue points, \texttt{locator} models are trained and evaluated on datasets with given S/N values separately. In orange points, \texttt{locator} is trained on the dataset with the lowest S/N and then applied to all datasets. In green points, \texttt{locator} is trained on the dataset with the highest S/N and then applied to all datasets.} }
    \label{fig:noise}
\end{figure}

As shown in the left panel of Figure~\ref{fig:t0te}, the scattering in $\lg t_{\rm E}$ increases for shorter events.
This is probably because, as timescale decreases, fewer data points are available to \texttt{locator} in the parameter inference. For example, with $k=1/3$ there are only $\sim25$ time stamps fall within $\Delta T$ for the shortest ($t_{\rm E}=5\,$days) event.
To improve the performance of \texttt{locator} on short events, we select light curves with $t_{\rm E}<40\,$days and train another model with $k=2$. After 8 epochs of training with the same learning rate schedule, the loss drops to 0.083 on the validation set. We evaluate the new model on 16,384 light curves with $t_{\rm E}<40\,$days.
\footnote{This includes light curves used in Figure \ref{fig:t0te} with $t_E<40$ and additional ones in the test set that add up to 16,384.}
The results are shown in Figure~\ref{fig:t0te_small}. The prediction accuracy on the timescale increases by a factor of $\sim2$, reaching a fractional uncertainty on $t_{\rm E}$ of $6.2\%$. The performance on the ratio $t_0/t_{\rm E}$ remains largely unchanged.

The above investigation has two implications. First, it outlines a general strategy to obtain more accurate $t_{\rm E}$. For example, 
one may divide the overall range of $t_{\rm E}$ into multiple smaller intervals and train separate \texttt{locator} models.
The preliminary estimate of $t_{\rm E}$ from the default run of \texttt{locator} can then be ``polished'' by the \texttt{locator} model that is optimized for the narrower $t_{\rm E}$ interval including the preliminary estimate.
Additionally, the comparison in \texttt{locator} performance between two $k$ values also sheds light on how \texttt{locator} works. In the single-lens case, the most constraining power on $t_0$ comes from the central region with width of $\sim u_0 t_{\rm E}$, whereas the event timescale $t_{\rm E}$ is best constrained by data points extending into the light curve wings. This seems to apply to \texttt{locator} as well. As shown in Figure~\ref{fig:rmse_k}, the timescale parameter can be better constrained when a wider mask window (i.e., a larger $k$) is used, whereas the event peak time $t_0$ is largely unaffected by the choice of $k$. More works are needed to better understand the mechanism of \texttt{locator} (and U-Net in general).

{We have fixed the magnitude of the baseline object at 18 and used S/N=33 for all data points in all simulated events. In reality, the baseline magnitude varies from event to event, and the photometric uncertainty varies from data point to data point. However, as discussed previously, the choice of the baseline magnitude does not really matter to the performance of \texttt{locator}, as long as the photometric observations cover a long enough time baseline. We therefore only focus on the impact of varying S/N on the performance of \texttt{locator}. Rather than simulating the realistic noise conditions,}
\footnote{{In real applications, one can and should preprocess the original light curve via, for example, outlier removal and data point rebinning, to make its data quality similar to that of simulated light curves.}}
{we study the impact of different training strategies. Several datasets with different S/N values are generated following the method in Section~\ref{sec:unet_data}. We then train \texttt{locator} models on light curves with a fixed S/N value and then evaluate the performance on light curves from the same dataset. The result of this training strategy is shown as the blue points in Figure~\ref{fig:noise}. Here we only show the RMSE values of $t_0$ as the trend is very similar for $t_{\rm E}$. Next, we train one \texttt{locator} model on events with the lowest quality (S/N=10) and apply it to all datasets, the result of which is shown in orange in Figure~\ref{fig:noise}. In the third test, we train \texttt{locator} on events with the highest quality (S/N=1000) and apply it to all datasets. This result is shown in green in the same figure. As expected, \texttt{locator} performs the best if it is trained and evaluated on light curves with the same (or very similar) data quality. However, the performance of \texttt{locator} is only slightly worse if it is trained on the noisest dataset. This is probably because \texttt{locator}, similar to other convolutional neural networks \citep{goodfellow2016deep}, is effectively composed of a series of filters learned from the data. When faced with noisy data, it learns to denoise through training, so the relatively large noises in the training data make \texttt{locator} more robust. For the same reason, the accuracy of the predicted $t_0$ is substantially worse on noisy light curves if \texttt{locator} only saw during the training light curves of the highest quality. For applications to real events, it seems that training one \texttt{locator} on relatively noisy data may be a good strategy that balances efficiency and accuracy.}

Finally, we discuss the issue related to wide ($s>1$) binaries. As pointed out earlier, the match between the ground truth labels and the predictions of \texttt{locator} for wide binaries is not as good as for close binaries. This larger mismatch, however, does not mean that the inference power of \texttt{locator} is reduced for wide binaries. After visually inspecting a large number of wide binary light curves, we find that the mismatch is related to the use of the reference system. When generating the light curves, we have followed the common convention and adopted the center of mass of the binary as the origin of the coordinate system. However, for wide binaries, the external shear due to a component at wide separation is better modeled in the reference frame defined on the center of magnification. For $s>1$, the offset between the center of mass and the closer center of magnification is \citep[e.g.,][]{DiStefano:1996, Chung:2005,penny2014speeding}
\begin{equation}
    \Delta x = \frac{1}{1+q} \left( s-\frac{1}{s}  \right) \cdot \left\{ \begin{array}{ll}
        q, & {\rm if~primary} \\
        1, & {\rm if~secondary} \\
    \end{array} \right. .
\end{equation}
For very wide binaries, the event timescale given by the mass of the closest lens also takes over the timescale given by the total mass of the binary. Both can lead to large offsets between the ground truth window and the predicted window. The most extreme example that has the largest deviation in $t_0$ and $t_E$ with $q=0.99$ and $s=2.8$ is shown in Figure~\ref{fig:lc_s_wide}. Once the reference frame defined on the center of magnification is adopted, the match between the ground truth and the prediction of \texttt{locator} is substantially improved (see also \citealt{zhang2021real}).

%% file: sec_ncde.tex
\subsection{Method} \label{sec:ncde_method}

We develop a separate method to infer the time-independent microlensing parameters, which we re-parameterize as
\begin{equation}
    \omega \equiv (u_0, \lg{q}, \lg{s}, \alpha/180^\circ, \lg{f_{\rm S}}) .
\end{equation}
As shown in Figure~\ref{fig:model}, this \texttt{estimator} consists of two main parts. A neural controlled differential equation (neural CDE) is used to extract features from the microlensing light curve, which is irregularly sampled and may contain large data gaps. The second part, mixture density network (MDN), infers the probability distribution of the microlensing parameters.

To improve the training speed and reduce the memory requirement, we first apply the log-signature transform \citep{morrill2021neural}. This transform outputs a shorter, steadier, and higher dimensional sequence that summarizes the sub-step information. For the simulated light curves with 500 time stamps (see Section~\ref{sec:ncde_data}), we set the log-signature depth $n_D=3$ and the shortened length $l=100$. The output thus has $v=5$ dimensions.

Neural CDE approximates the underlying process of the time series as a differential equation \citep{kidger2020neural,kidger_thesis}
\begin{equation} \label{eqn:ncde}
    z_{t} = \xi_{\theta} (X_{t_0}) +\int_{t_{0}}^{t} f_{\theta}\left(z_{t'}\right) \mathrm{d} X_{t'} .
\end{equation}
Here the latent state $z_t$ is defined on the time interval of the input signal, $(t_0, t_n)$, with $w=32$ dimensions.
The continuous signal $X_t$, defined on the same time interval, is obtained from applying natural cubic splines to the input, discrete signal. 
\footnote{{
Note that the use of cubic splines does not mean that neural CDE is just another interpolation scheme. This is solely for creating a continuous embedding of the discrete input. In contrast, standard interpolation schemes only make use of the values at fixed time stamps. A detailed discussion on this issue can be found in \cite{kidger_thesis}.}}
The initial value and signal-derivative of the latent state, $\xi_{\theta}$ and $f_{\theta}$, are two different neural networks, where the subscript $\theta$ denotes the dependence on the learnable machine learning parameters. By Equation~\ref{eqn:ncde}, neural CDE gradually extracts features from the signal $X_t$ according to a learned policy $f_\theta$, records its knowledge in the latent state $z_t$, and output a terminal value $z_{t_n}$ that summarizes the useful information of the input signal. We refer interested readers to \citet{kidger2020neural} and \citet{kidger_thesis} for further details of neural CDE.

The summary features from neural CDE is then fed into the MDN \citep{bishop2006pattern} to estimate the probability distribution of the microlensing parameters
\begin{equation} \label{eqn:mdn}
    p(\omega|z_{t_n}) = \sum_{i=1}^{n_{\rm G}} \pi_i(z_{t_n}) \phi_{\bar{\omega}_i(z_{t_n}), \Sigma_i(z_{t_n})}(\omega).
\end{equation}
Here $\phi_{\bar{\omega}, \Sigma}$ denotes the density of a multivariate Gaussian with mean $\bar{\omega}$ and covariance matrix $\Sigma$. The normalized weight $\pi_i$, mean $\bar{\omega}_i$, and covariance matrix $\Sigma_i$ of each of the $n_{\rm G}$ multivariate Gaussians are given by the neural network of MDN.
Compared to the masked autoregressive flow method of \citet{zhang2021real}, MDN provides a straightforward and efficient alternative with explicit density estimation, which is useful for the identification of degenerate solutions and the further optimization (see below). 
We adopt $n_{\rm G}=12$ Gaussians with diagonal covariance matrices. 
{This preserves universality as long as $n_G$ is large enough \citep{bishop2006pattern}.}

In the joint framework, \texttt{estimator}, the initial value network $\xi_\theta$, the evolution network $f_\theta$, and MDN are constructed as residual networks (ResNet, \citealt{he2016deep}) with three residual blocks and each block consisting of two fully connected layers of width 1024. The evolution network has an additional activation of hyperbolic tangent to rectify the dynamic. To train the model, we minimize the negative log-likelihood (NLL) loss function
\begin{equation}
    \mathcal{L}_{\text{estimator}} = \mathop{\mathbb{E}}\limits
    _{{\text{light curve }}} [-\ln p(\omega_{\rm true} | z_{t_n})],
\end{equation}
To further optimize the microlensing parameters, we start from mean positions of the multivariate Gaussians and minimize the light curve $\chi^2$ values via the \citet{Nelder:1965} simplex algorithm. In this step, the source flux parameter $f_{\rm S}$ is derived analytically for a given set of $(q, s, u_0, \alpha)$ by maximizing the model likelihood \citep[e.g.,][]{mulensmodel}.

\begin{figure}
    \centering
    \includegraphics[width=\linewidth]{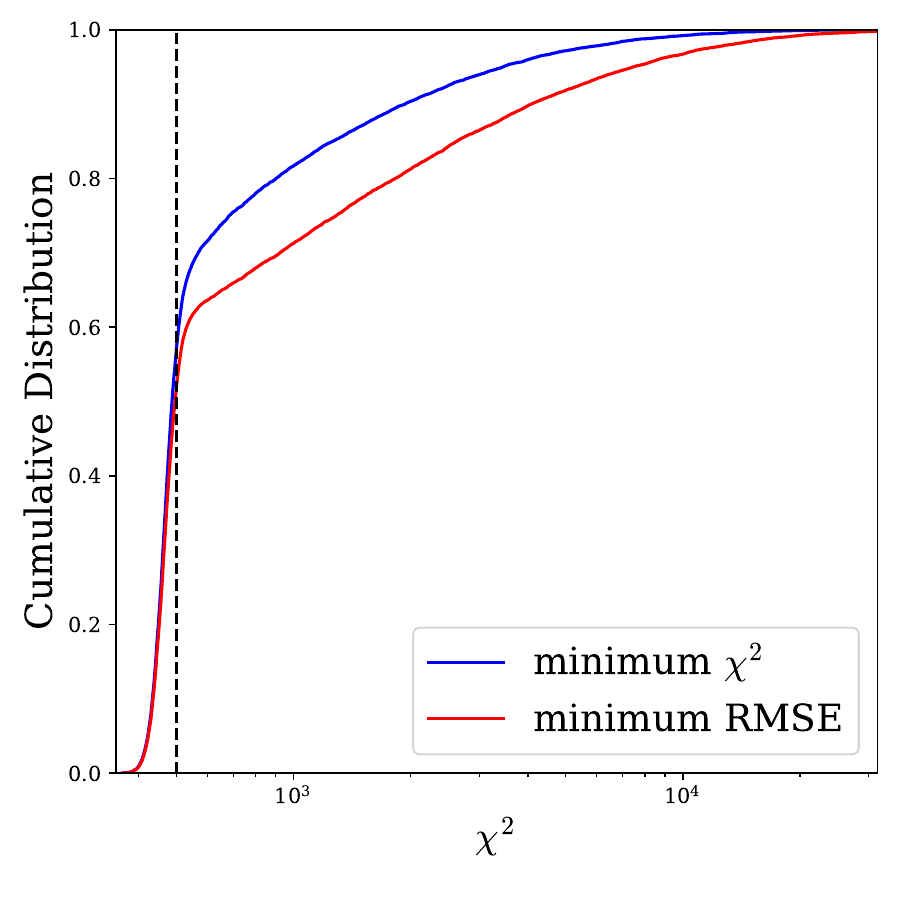}
    \caption{Cumulative distributions of the $\chi^2$ values of the output microlensing models of \texttt{estimator}. In blue and red curves models with the minimum $\chi^2$ value and the closest distance (in parameter space) to the input model are used, respectively. The vertical dashed line indicates $\chi^2=500$, which is approximately the medians for both curves.
    \label{fig:cdf_chi2}}
\end{figure}

\begin{figure}
    \centering
    \includegraphics[width=\linewidth]{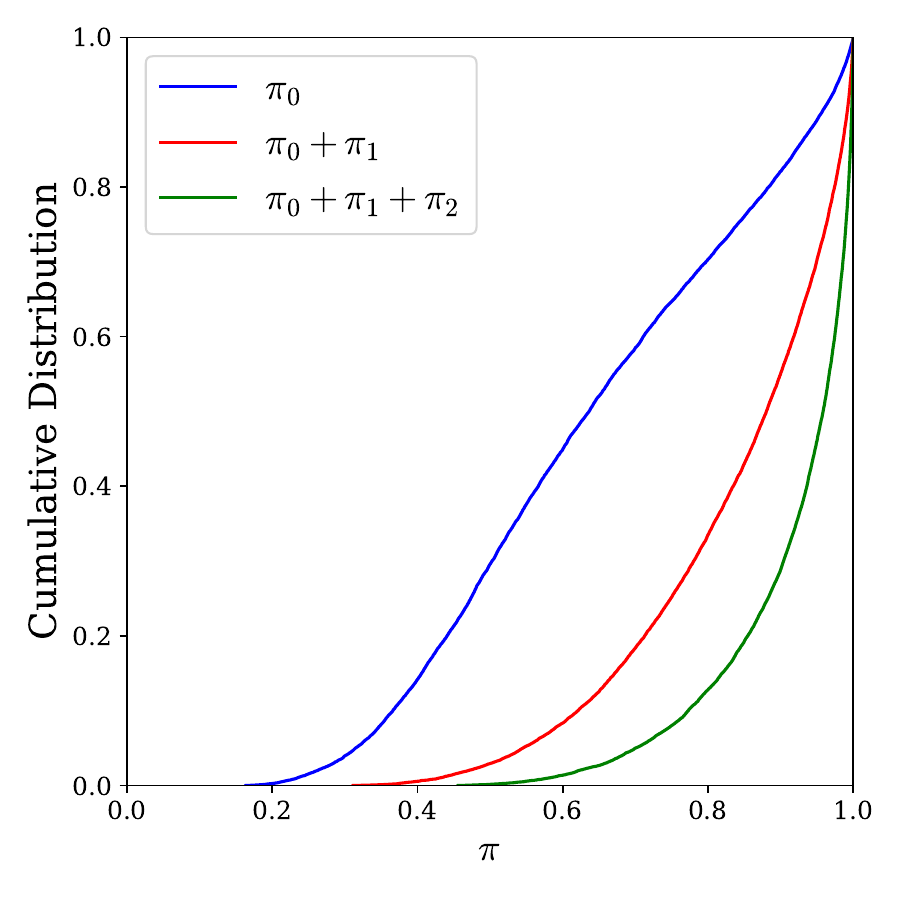}
    \caption{Cumulative distributions of the normalized weights in MDN. Here $\pi_0$, $\pi_1$, and $\pi_2$ are the top three highest weights in our MDN of 12 multivariate Gaussians. The blue curve indicates that there is one dominating Gaussian in about half of the cases, and the green curve shows that the combination of the top three Gaussians dominates the density distribution in almost all cases.} 
    \label{fig:cdf_pi}
\end{figure}

\begin{figure*}
    \centering
    \includegraphics[width=\linewidth]{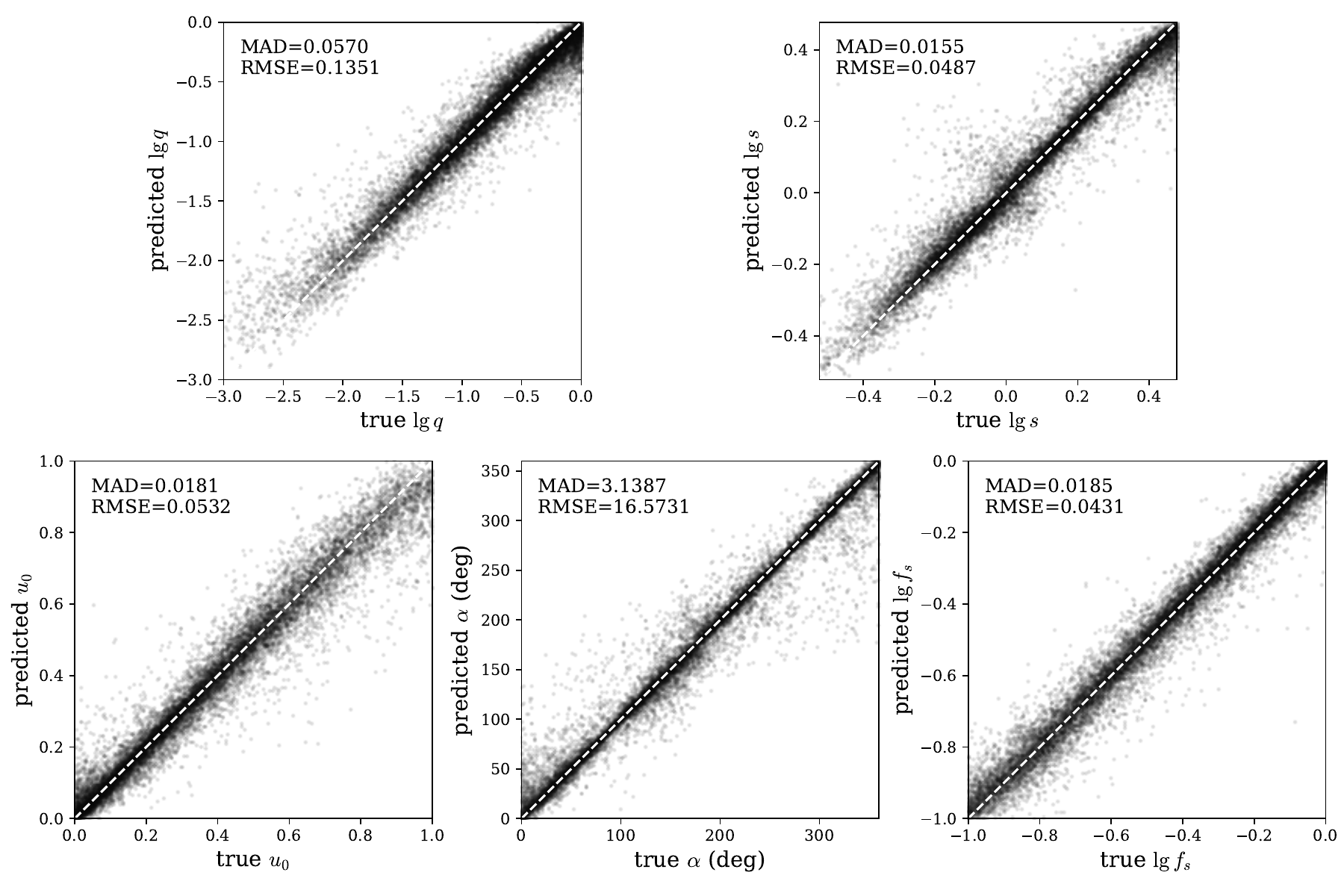}
    \caption{Comparisons between the input parameter values and the direct output of \texttt{estimator}. Out of the 12 Gaussian means of MDN, we identify the one that is closest to the ground truth in the parameter space and take it as the prediction. The median absolute deviation (MAD) is used to evaluate the agreement, and the root-mean-squared error (RMSE) is also provided for a reference.
    \label{fig:mdn_rmse}}
\end{figure*}

\begin{figure*}
    \centering
    \includegraphics[width=\linewidth]{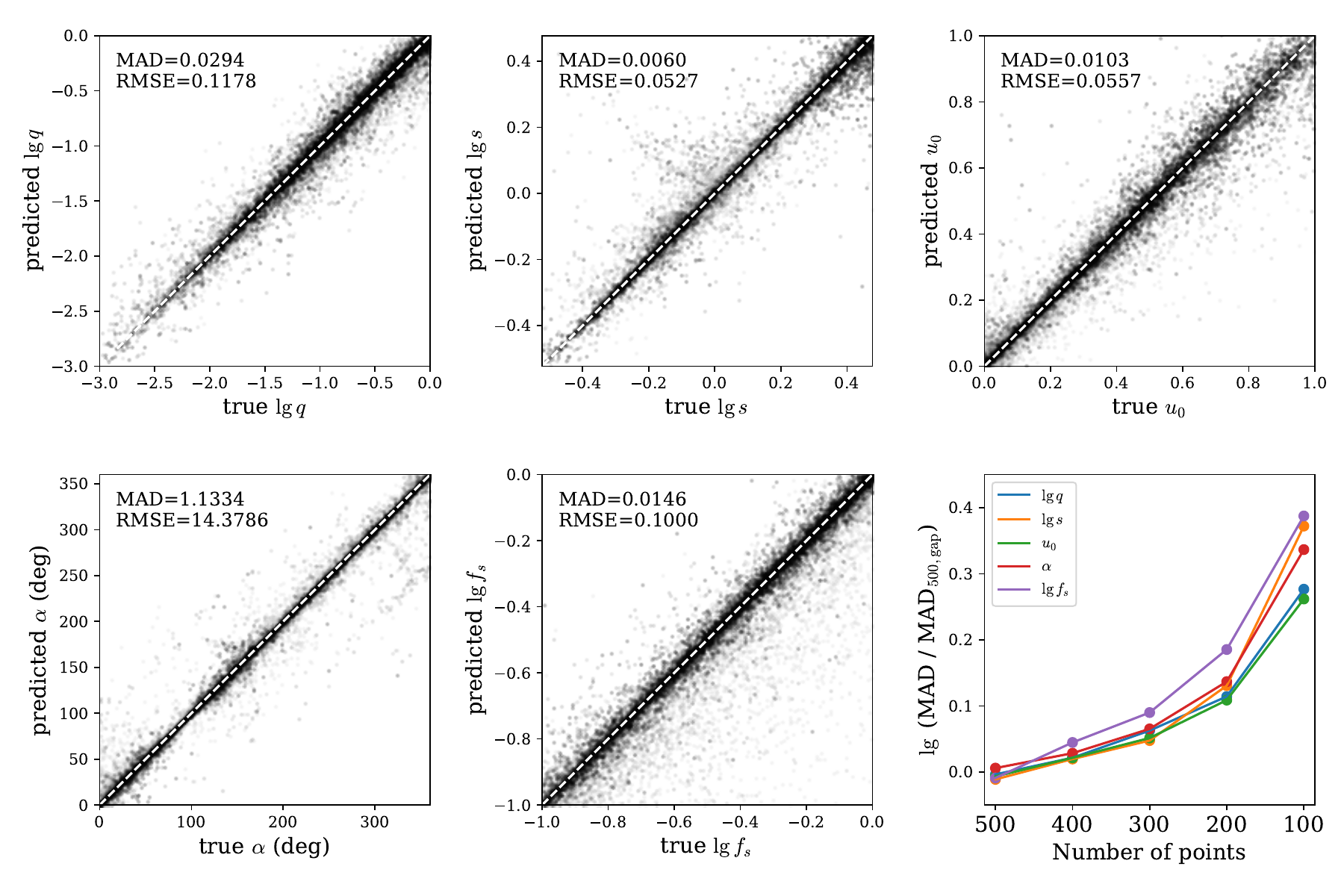}
    \caption{All except the lower right panel show comparisons between the input and the optimized parameter values. From the means of the 12 multivariate Gaussians, we apply \citet{Nelder:1965} simplex method and obtain optimized model parameters. The set of parameters closest to the ground truth in the parameter space is used in these comparisons. The lower right panel shows the prediction accuracy of microlensing parameters evaluated on events with various numbers of data points and no gap. Here MAD$_{500,\rm gap}$ is the value of MAD evaluated on events with 500 points and data gap, as incidated in the top left corner of the other panels.
    \label{fig:opt_close}}
\end{figure*}

\begin{figure*}
    \centering
    \includegraphics[width=\linewidth]{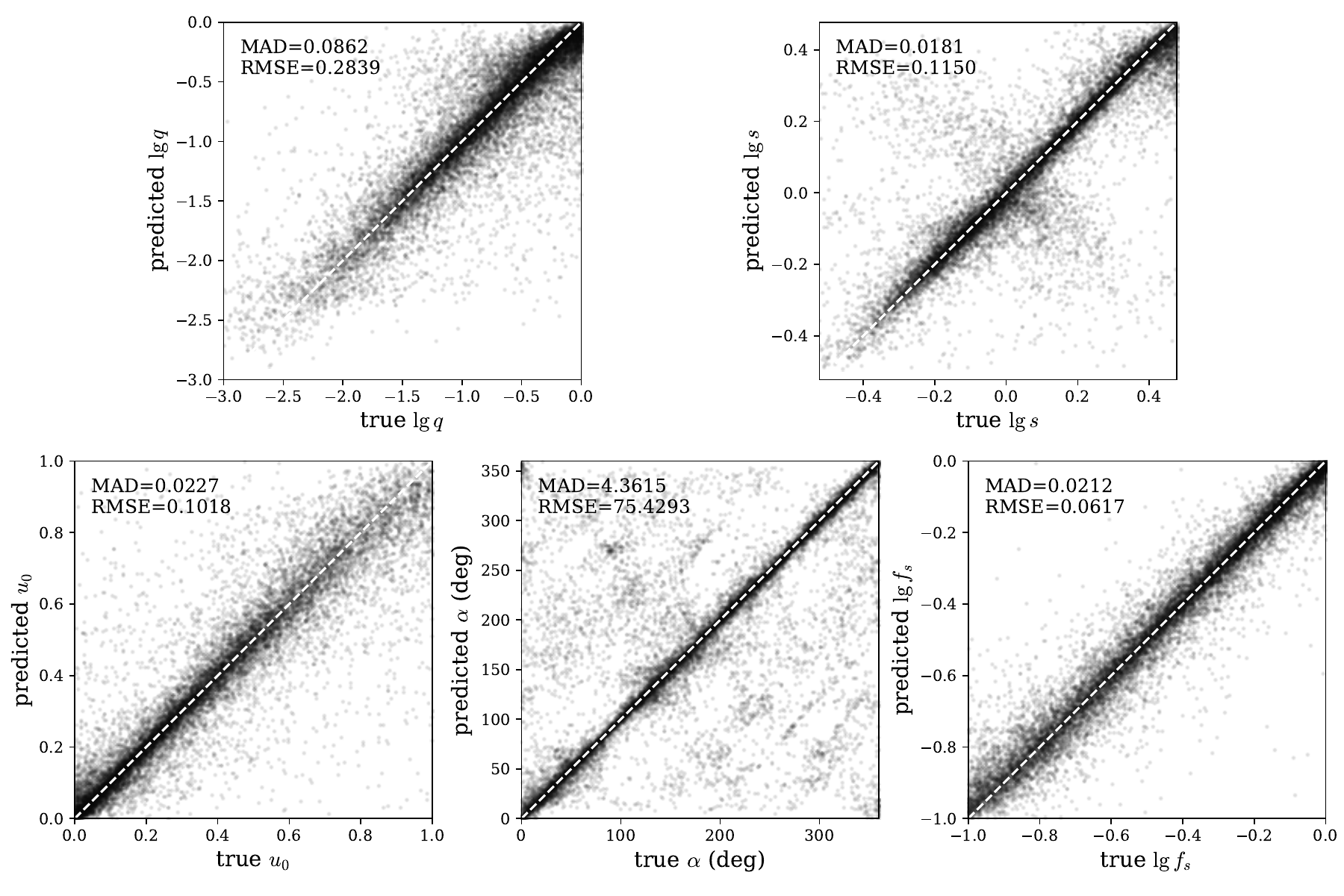}
    \caption{{Similar to Figure \ref{fig:mdn_rmse} except that the Gaussian with the largest weight is taken as the prediction.}
    \label{fig:mdn_pi}}
\end{figure*}

\begin{figure*}
    \centering
    \includegraphics[width=\linewidth]{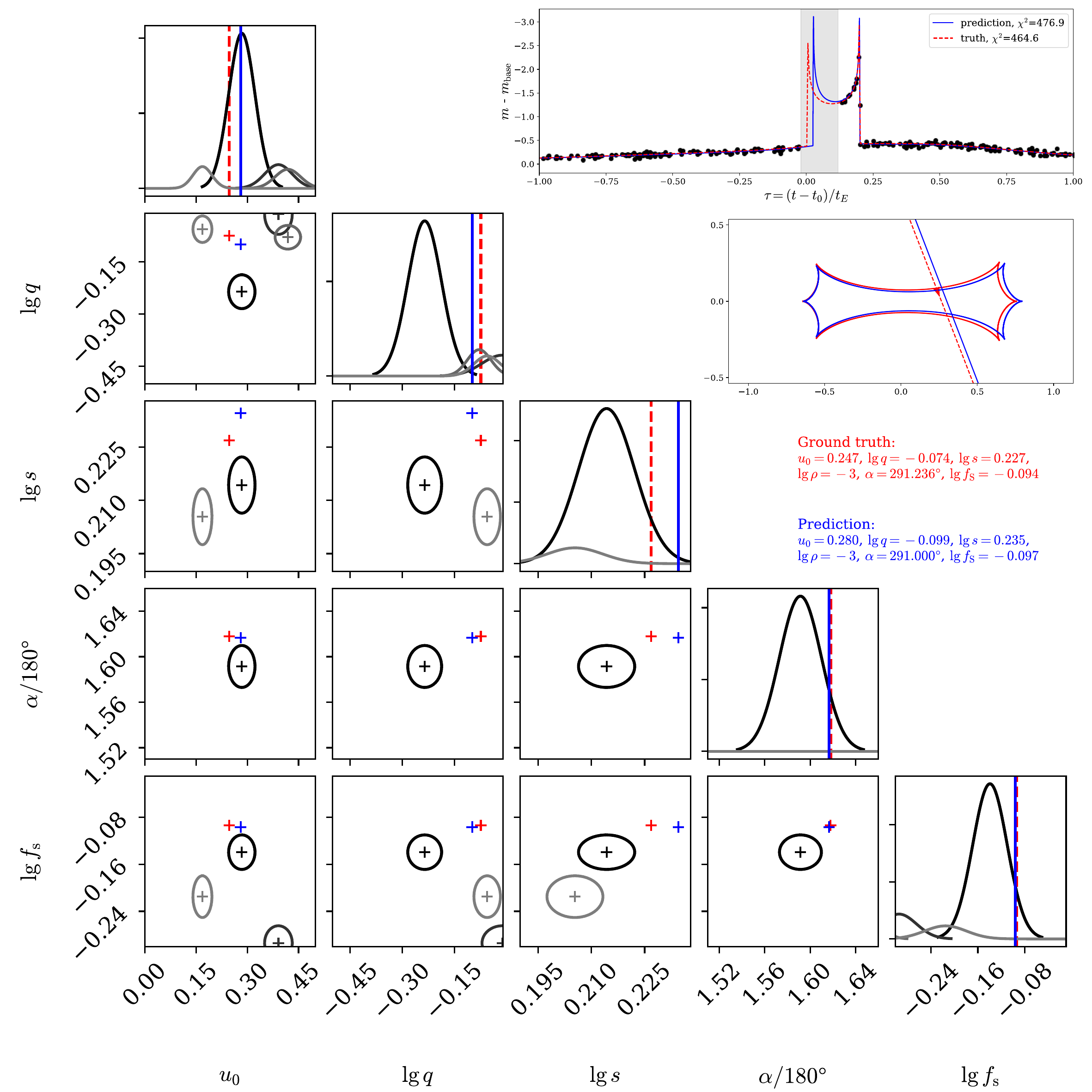}
    \caption{An example event in which the binary caustic-crossing feature is partially missing due to the injected data gap, as shown by the light curve in the top right corner. Red color is for the input model, whereas blue color is for the prediction given by \texttt{estimator}. The corner plot shows the probability distributions of the four MDN Gaussians with the highest weights, with the colors indicating the relative weights. Contours denote the 1-$\sigma$ boundaries of the multivariate Gaussians with diagonal covariance matrices. As we have chosen axis limits to optimize the overall appearance, not all four contours are visible in every panel. Values and positions of the ground truth and the prediction of \texttt{estimator} on the corner plot are also indicated. Even without the final optimization, the Gaussian distribution with the highest weight gives a fairly good approximation of the input parameters.}
    \label{fig:example_gap}
\end{figure*}

\subsection{Data generation} \label{sec:ncde_data}

For the purpose of learning the time-independent parameters, we generate light curves as functions of the parameter $\tau$. This is equivalent as setting $t_0=0$ and $t_{\rm E}=1\,$d. The other parameters are sampled in the same way as in Secion~\ref{sec:unet_data}.

Each light curve contains 500 data points randomly drawn within the interval $-2 \leq \tau \leq 2$. For a typical timescale $t_{\rm E}=20\,$d, this corresponds to an averaged cadence of $\sim$4\,hr
{with a range from 24\,min (lower 10\%) to 8.8\,hr (upper 10\%).}
To mimic the missing data points due to bad weather and/or instrumental failures, we throw away data points within one randomly chosen light curve segment out of the 25 equal-duration segments that constitute the whole light curve. This leaves a data gap of $3.2\,$d for the aforementioned timescale. Light curves and noises are generated in the same way as in Section~\ref{sec:unet_data}. Finally, we only keep light curves with significant binary features, defined as $\chi_{\rm pspl}^2$/d.o.f.\ $>2$.

Six data sets are generated, each having $10^5$ light curves. We use five of them for training \texttt{estimator} and the last one for validation and test. 
To further understand the performance of \texttt{estimator} on light curves with different cadences, we have also generated data sets with various numbers of data points within the interval $-2 \leq \tau \leq 2$. For these additional data sets, we have assumed irregular sampling but no gap.

\subsection{Performance of \texttt{estimator}} \label{sec:ncde_perform}

We train \texttt{estimator} with the ADAM optimizer \citep{kingma2014adam}. The learning rate, initially set to $10^{-4}$, drops by ten percent after each epoch and reaches a minimum of $10^{-6}$. Training samples are labeled by their input parameters $\omega$ and divided into mini-batches of size 128. The training took $\sim1\,$day for 50 epochs on one NVIDIA Tesla V100 GPU, and the average loss drops to $-8.4$ on the validation set of size 1024. 

We apply the optimized \texttt{estimator} to 16,384 light curves from the test set and obtain the probability distribution of the microlensing parameters given by the summation of 12 multivariate Gaussians. We then apply the \citet{Nelder:1965} simplex algorithm to get the refined microlensing parameters. 
We show in Figure~\ref{fig:cdf_chi2} the cumulative distributions of the model $\chi^2$ values of two selected sets of parameters, one that has the smallest $\chi^2$ value and the other that is closest to the input position in the parameter space. The median $\chi^2$ values are $\sim500$ for both solutions, comparable to the number of data points in the simulated light curve and thus approximately the degrees of freedom. This suggests that the found solutions can indeed describe the data very well. Even though in $\sim$20\%--30\% of cases the model $\chi^2$ values are substantially large compared to the number of data points, the returned parameters are still good representatives of the input parameters, as will be discussed later.
{Appendix~\ref{app:param_vs_lc} shows an example event whose model prediction gives large $\chi^2$ and yet reasonably accurate binary parameters.}

We have included a mixture of $n_{\rm G}=12$ Gaussians in the density estimate. Is it flexible enough? To answer this question, we first look at the distribution of the Gaussian weights given by \texttt{estimator}. As shown in Figure~\ref{fig:cdf_pi}, the probability distribution is dominated by one Gaussian in $\sim$50\% of the tested events and by the top three Gaussians with the highest weights in nearly all tested events. As for the optimized solutions, 74\% and 85\% of the minimum $\chi^2$ and minimum RMSE solutions are given by the top four Gaussians, respectively. Only in $\lesssim1\%$ of the tested events does the Gaussian with the lowest weight lead to either solution. Therefore, we conclude that 12 Gaussians can satisfactorily describe the probability distribution of microlensing model parameters.

We compare the \texttt{estimator} predictions before and after the simplex optimization with the input parameters in Figures~\ref{fig:mdn_rmse} and \ref{fig:opt_close}. 
For each tested event, we choose as the model prediction the set of parameters that is closest to the input position in the parameter space, in order to minimize the impact of degenerate solutions. 
{The results change only marginally if we pick the Gaussian with the highest weight as the model prediction, as seen in Figure~\ref{fig:mdn_pi}.}
We use the median absolute deviation (MAD) to evaluate the agreement between input and prediction. Compared to RMSE, MAD is more robust against outliers.
As shown in Figure~\ref{fig:mdn_rmse}, the direct output of \texttt{estimator} already contains a set of parameters that is fairly close to the input truth.
Among all parameters, binary mass ratio $q$ and separation $s$ are of special interest to the physical interpretation of the event. The direct output of \texttt{estimator} yields a typical deviation of $0.057$ ($0.0155$) for $\lg{q}$ ($\lg{s}$), corresponding to a fractional uncertainty of $13\%$ ($3.6\%$) on $q$ ($s$).
The accuracy of the prediction is further improved after the optimization, as shown in Figure~\ref{fig:opt_close}. Specifically, the optimized prediction can achieve median deviations of $\sim0.03$ and $\sim0.006$ for $\lg{q}$ and $\lg{s}$, corresponding to fractional errors of $\sim7\%$ and $\sim1.4\%$ in $q$ and $s$, respectively.
Overall, the agreement between the input and predicted values is very well, demonstrating the high accuracy of our method on light curves with realistic sampling conditions.

We have also applied the optimized \texttt{estimator} to light curves in the additional data sets, which have different averaged cadence but no data gap.
\footnote{That is, we directly apply \texttt{estimator} without retraining.}
The results are shown in the lower right panel of Figure~\ref{fig:opt_close} for the optimized prediction. There are a few interesting trends to report. First, the performance of \texttt{estimator} is better, although fairly marginally, on events with 500 points and no data gap. Second, the prediction accuracy decreases with decreasing number of data points as expected, but the degradation is typically small. Specifically, with 200 data points the prediction accuracy is only worse by $\sim10^{0.13}-1 \approx 35\%$ compared to the default case with 500 data points. The only exception is seen in the case with 100 data points, for which the performance is reduced by a factor of $\gtrsim 2$. This is probably because the averaged cadence becomes too low to constrain binaries with mass ratios down to $10^{-3}$.
\footnote{The averaged cadence in terms of $\tau$ is $4/100$, and thus the smallest binary mass ratio that it can detect is $\gtrsim (4/100)^2 = 1.6 \times 10^{-3}$.}
Nevertheless, the overall uncertainty in the case of 100 points remains fairly small and acceptable. These results may provide useful guidance to the future application of our method to real light curves with variant cadences and/or the detection of low mass-ratio planets.

We show in Figure~\ref{fig:example_gap} an example event whose anomalous feature is only partially covered by data points. The probability distribution from \texttt{estimator}, represented by the multivariate Gaussians, is already at the neighborhood of the ground truth position. The optimization with \citet{Nelder:1965} algorithm leads to even better match in microlensing parameters as well as model the light curve. Events like this one, which have irregular samplings and large data gaps, pose challenges to machine learning methods that base on the standard RNN approach.

Our method also identify multiple solutions in events that suffer model degeneracy. We defer to Section~\ref{sec:0371} for further discussions on this issue.

%% file: sec_application.tex
\subsection{Joint pipeline} \label{sec:joint}

Combining the methods described in Sections~\ref{sec:unet} and \ref{sec:ncde}, the joint pipeline of MAGIC works as follows. First, we feed the light curve into \texttt{locator}, which outputs the predicted $t_0$ and $t_{\rm E}$. If $t_{\rm E}$ is small, we can further refine it using \texttt{locator} with a larger $k$. Then we use the predicted $t_0$ and $t_E$ to shift and rescale the light curve. The resulting light curve is processed by \texttt{estimator}, which in return gives the positions and shapes of $n_G=12$ Gaussians in the parameter space. If needed, these Gaussians can be further optimized to obtain more accurate prediction or predictions (see Section~\ref{sec:ncde_method}).

The performance of the combined pipeline is largely determined by the propagation of inaccurate parameters from \texttt{locator} to \texttt{estimator}. We therefore test the joint pipeline on 16,384 light curves from the test set of \texttt{locator}. After applying \texttt{locator}, we transform the light curves using the predicted values of $t_0$ and $t_{\rm E}$. The transformed light curves are then sent to \texttt{estimator} to infer the time-independent microlensing parameters. 
The set of parameters closest to the position of the input values in the parameter space is adopted as the final prediction.

\begin{figure*}
    \centering
    \includegraphics[width=\linewidth]{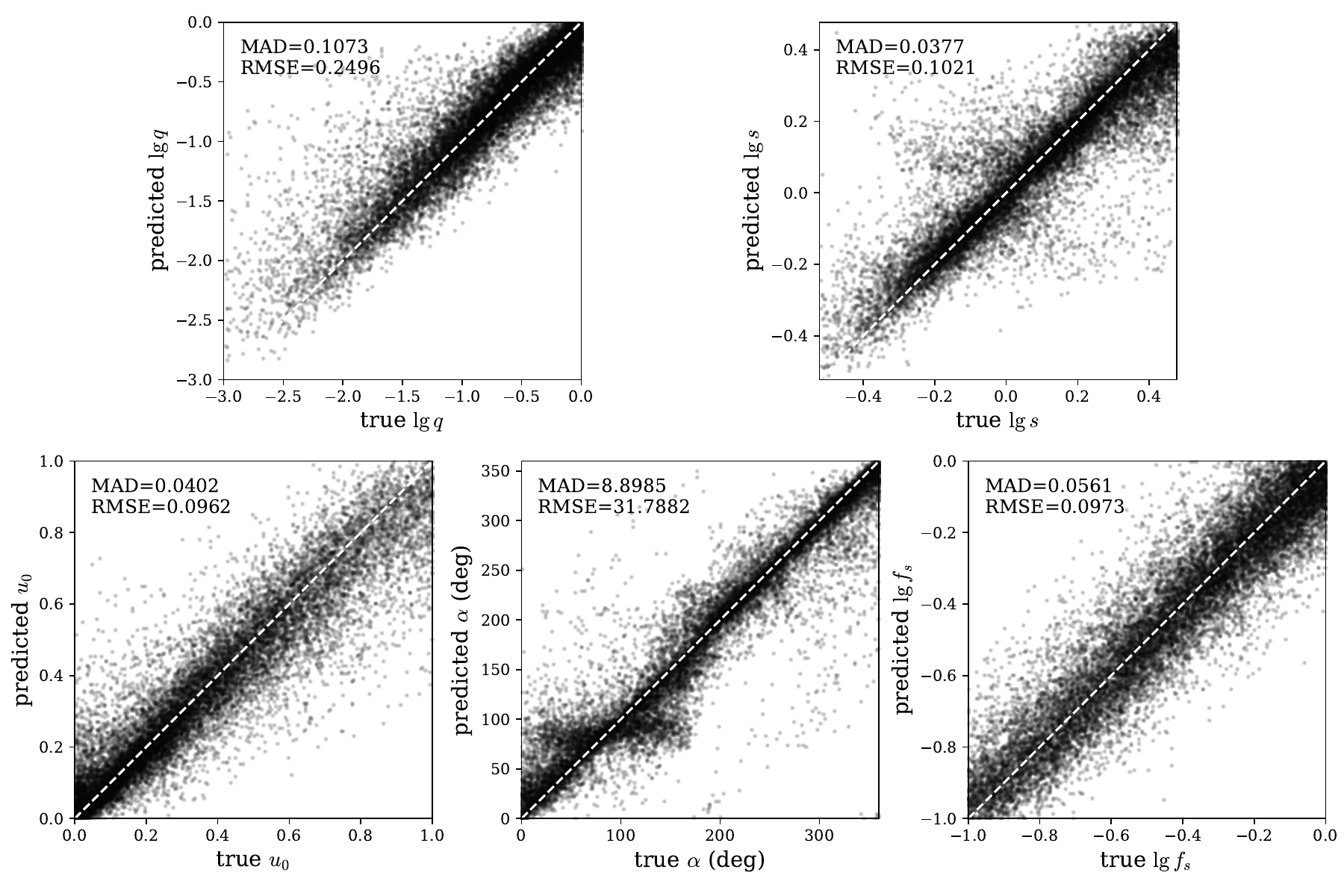}
    \caption{Performance of the joint pipeline. Simulated light curves for \texttt{locator} are used in this test. After the application of \texttt{locator}, the light curve is sent to \texttt{estimator} for the inference of the time-independent parameters.} 
    \label{fig:joint_close}
\end{figure*}

The comparisons between the input and the predicted values of the time-independent parameters are shown in Figure~\ref{fig:joint_close}. The typical, fractional deviation in $q$ and $s$ are $25\%$ and $9\%$, respectively. These deviations remain small, even though they are worse than the \texttt{estimator}-only case by a factor of $\sim2$. 
Overall, the joint pipeline achieves fairly high prediction accuracy in the time-independent parameters even with the use of the inaccurate time-dependent parameters.
We expect the accuracy to be further improved once the predictions are optimized via the \citet{Nelder:1965} algorithm, as seen in the \texttt{estimator}-only case.

A substantial fraction of binary events can be well described by a \citet{Paczynski:1986} model plus relatively short binary perturbation. For such events, one may mask out the anomalous feature and fit a single-lens model to obtain precise and accurate peak time $t_0$ and event timescale $t_{\rm E}$. Combining these values with \texttt{estimator}, we can achieve nearly as accurate time-independent parameters as we do in the \texttt{estimator}-alone case.

It is worth noting that the light curves generated for \texttt{locator} are similar to but not the same as those used for training \texttt{estimator}. 
First, \texttt{locator} uses binary light curves with $\chi^2_{\rm pspl}/{\rm d.o.f.}>1.25$, whereas \texttt{estimator} uses binaries with $\chi^2_{\rm pspl}/{\rm d.o.f.}>2$. Both correspond to a $\Delta \chi^2>500$ between single and binary models, under the reasonable assumption that the binary signals are usually local features. 
\footnote{Although we have focused on such binaries with strong signals, we expect our method to work on weak binaries as well.}
Furthermore, the time window is fixed and the timescale parameters varies in the data set for \texttt{locator}. This means the number of data points within $-2 \leq \tau \leq 2$ varies too. Nevertheless, we have shown in Section~\ref{sec:ncde_perform} that the performance of \texttt{estimator} is only mildly affected with the various length of input time series.

In addition to its high accuracy, our method demonstrates high efficiency in the parameter estimation. Specifically, \texttt{locator} and \texttt{estimator} can process $\sim1000$ and $\sim400$ light curves per second on a single GPU, respectively. As shown in Figure~\ref{fig:opt_close}, the output of \texttt{estimator} is already accurate enough for purposes like statistical studies. If needed, the optimization step can be performed. Compared to the two machine learning steps, this traditional optimization step is slower, but there are only up to 12 initial positions to be optimized, which altogether may require up to a few thousand light curve evaluations. For comparison, the traditional approach usually requires optimizations at individual grid points on a fine grid in the $(\lg{q}, \lg{s}, \alpha)$ space \citep[e.g.,][]{Albrow:2000, Wang:2022}. Our approach therefore represents a speedup by factors of $\sim10^2$--$10^5$.

\subsection{A Real Event Application} \label{sec:0371}

We apply MAGIC to KMT-2019-BLG-0371 (hereafter KB190371 for short), which is a short-timescale ($t_{\rm E}=6.5\,$d) binary event with $q \approx 0.08$ or $0.12$ \citep{Kim:2021}. KB190371 is chosen for this demonstration based on several considerations. First, its caustic-crossing feature is prominent and U-shaped, which is typical for binary events. Second, this event suffers from the close/wide degeneracy \citep{Griest:1998, Dominik:1999}
\footnote{We use this term to keep consistent with the original discovery paper, even though recent studies have suggested that such ``close/wide degeneracy'' should be called offset degeneracy \citep{Zhang:2022}}
, thus enabling additional tests of MAGIC. Furthermore, except for the relatively short timescale, the other microlensing parameters are fairly typical in comparison to those used in our simulation.

The available data sets from \citet{Kim:2021} are not aligned to the same magnitude system. In order to align the data sets and then apply MAGIC, we mask out the caustic-crossing feature, fit a \citet{Paczynski:1986} model to the data, and use the best-fit flux parameters to align all other data sets to the OGLE one. We note that, although here we have made use of the special property of KB190371 to align the data sets, such a data alignment task is the same as deriving the source color and can be performed without reference to specific microlensing models \citep{Yoo:2004, Gould:2010a}, as has been practiced in the analysis of many real microlensing events \citep[e.g.,][]{Gould:2010, Zhu:2017}.

To apply our joint pipeline, we first preprocess the data and make the light curve similar to those used in the training procedure. To be more specific, we only include data points falling in the time interval of about 100 days centered at the anomaly, and filter out data points with large ($>0.1\,$mag) error bars. KB190371 was observed with high cadence, so we bin every five consecutive points into one. Finally, the light curve is shifted to the same base magnitude as those used in \texttt{locator}.

Then we feed the preprocessed light curve into \texttt{locator}. Because the timescale is relatively small, we set $k=2$. The output mask is rectified into binary values $\{0, 1\}$ with a threshold of $0.5$. This step gives the prediction $t_0~{(\rm HJD')}=8592.391$ and $t_{\rm E}=6.68\,$d, both of which are close to the values given in \citet{Kim:2021}. We transform the light curve with the predicted $t_0$ and $t_{\rm E}$ and only keep the segment within $-2 \leq \tau \leq 2$.

The rest of application follows closely the procedure given in Section~\ref{sec:joint}. With the positions of the 12 multivariate Gaussians given by \texttt{estimator}, we perform \citet{Nelder:1965} optimization to further refine the microlensing parameters. Because the light curve shows prominent finite-source effect, we also include in the optimization the scaled source size $\rho$.

\begin{figure*}
    \centering
    \includegraphics[width=\linewidth]{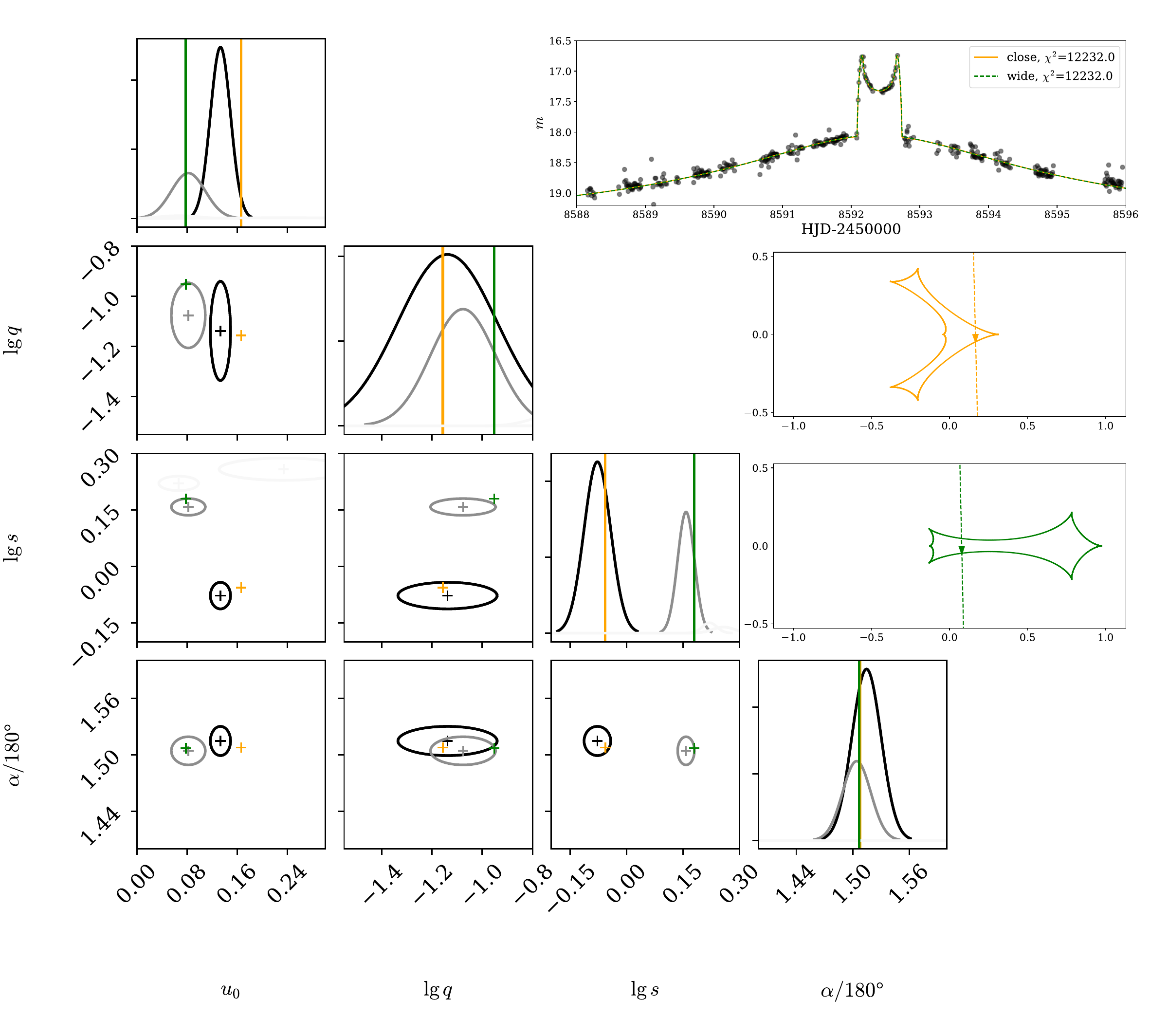}
    \caption{The application of MAGIC to KB190371 and the output. The light curve and the MAGIC predictions are shown in the top right panel. Two solutions, labeled ``close'' and ``wide,'' have comparable $\chi^2$ values, and the corresponding microlensing geometries (i.e., caustic curve and source trajectory) are shown below the light curve plot. The corner plot shows the positions of the optimized solutions as well as the top two Gaussians that correspond to the ``close'' and ``wide'' solutions.
    \label{fig:kmt}}
\end{figure*}

The results are shown in Figure \ref{fig:kmt}. The top two Gaussians with the highest weights match closely to the close ($s<1$) and wide ($s>1$) solutions, respectively. The corresponding model light curves from the optimized solutions both match the data reasonably well, with almost indistinguishable $\chi^2$ values. According to these solutions, the event can be explained by a binary with $(\lg{q}, \lg{s})=(-1.16, -0.057)$ or $(-0.95, 0.180)$. Given the imperfect alignment as well as the slightly different preprocessing procedures, we consider these values in good match with those given by the discovery paper.

\citet{Kim:2021} found the source scaled parameter $\rho$ to be $(6.46\pm0.17) \times 10^{-3}$ and $(6.99\pm0.21) \times 10^{-3}$ for the close and wide solutions, respectively. These are quite different from the value ($10^{-3}$) that we have used to train MAGIC. Nevertheless, the optimized parameters yield $\rho=6.6 \times 10^{-3}$ and $7.3 \times 10^{-3}$ for the close and wide solutions, respectively, in good agreement with the values from the discovery paper. In other words, the presence of the finite-source effect in KB190371 does not seem to confuse MAGIC, even though MAGIC has not learned to infer the source scaled parameter $\rho$. This is probably because the finite-source effect in KB190371 only affects a limited number of data points around the caustic entrance and exit. While the majority of binary microlensing events bear the same feature, some may have their binary signals substantially modified by the finite-source effect \citep[e.g.,][]{Hwang:2018}. Therefore, it may be necessary to include in MAGIC the ability to infer $\rho$. See Section~\ref{sec:disc} for further discussion.

\begin{figure*}
    \centering
    \includegraphics[width=\linewidth]{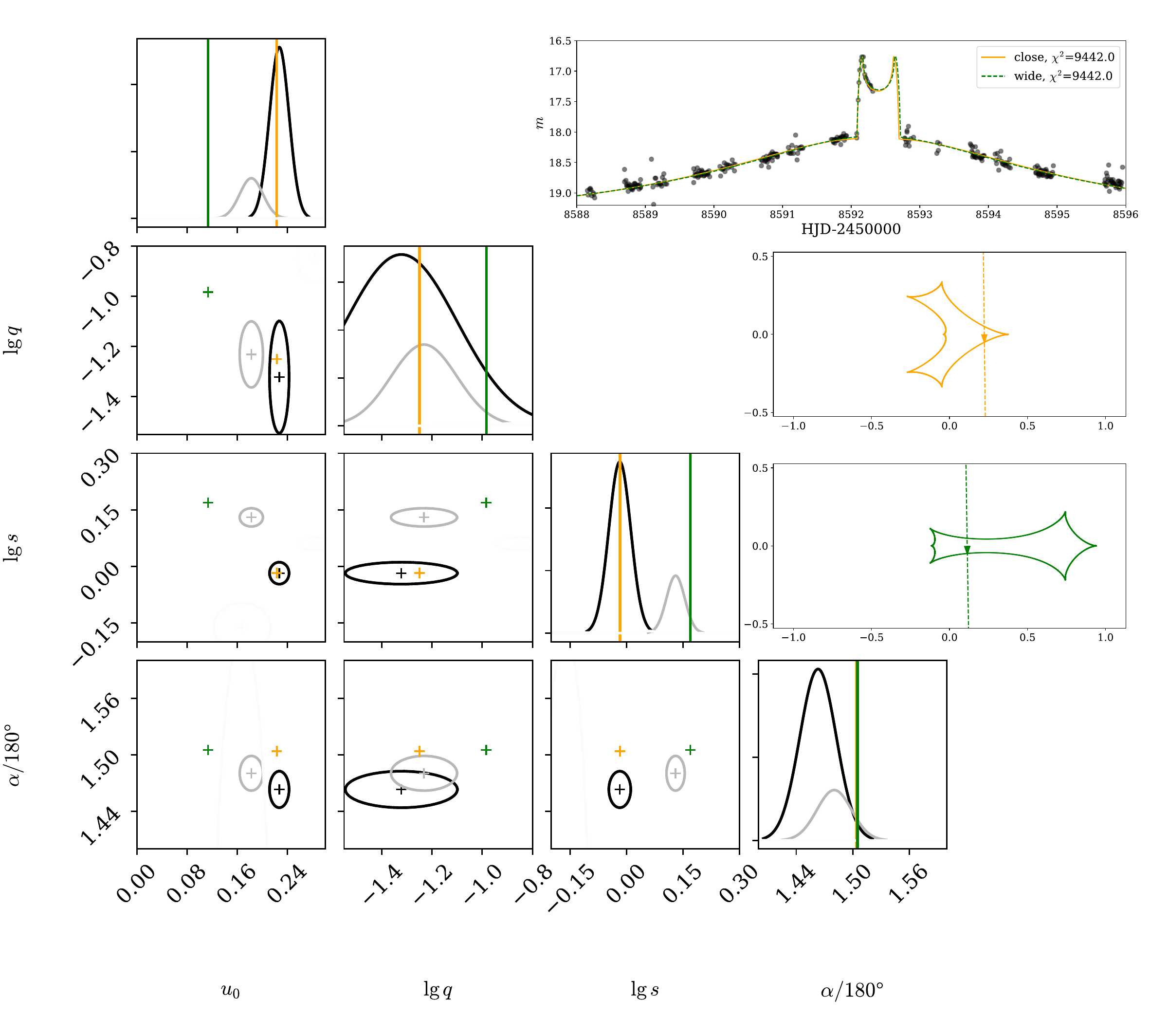}
    \caption{Similar to Figure~\ref{fig:kmt} except that observations from the South Africa site of KMTNet are excluded in this analysis. As a result, the exit portion of the caustic crossing feature is not covered with observations. Even so, MAGIC is able to identify the two degenerate solutions at high accuracy.
    \label{fig:kmt_gap}}
\end{figure*}

To further demonstrate the capability of MAGIC to handle large data gaps, we exclude observations from the South Africa site of KMTNet. This creates a substantial gap during the exit of the caustic crossing, as shown in the top right panel of Figure \ref{fig:kmt_gap}. The same procedure is applied as before, and the results are shown in Figure \ref{fig:kmt_gap}. The two degenerate solutions are again identified at high accuracy, indicating the robustness of MAGIC in handling large data gaps as well as model degeneracy.

%% file: sec_disc.tex
We present Microlensing Analysis Guided by Intelligent Computation (MAGIC), a machine learning method that is readily applicable to the analysis of binary microlensing light curves from ongoing and future photometric surveys. We divide the microlensing model parameters into two groups and develop separate deep learning neural networks. For the inference of time-dependent parameters, namely the peak time $t_0$ and the event timescale $t_{\rm E}$, we construct \texttt{locator}, which is a U-Net architecture \citep{ronneberger2015unet}. For the inference of time-independent parameters, namely $(u_0, q, s, \alpha, f_{\rm S})$, we construct \texttt{estimator}, which uses neural CDE \citep{kidger2020neural,kidger_thesis} to extract features and mixture density network (MDN; \citealt{bishop2006pattern}) to infer the probability distribution. The output parameters can be further refined through the standard optimization method. Depending on properties of the event under investigation, the two components can be used separately or jointly to achieve the best performance.

\textit{A little MAGIC can take you a long way}.
\footnote{From Roald Dahl.}
We train and validate both \texttt{locator} and \texttt{estimator} as well as the joint pipeline on simulated light curves. To mimic the sampling condition in the realistic world, we introduce irregular sampling and large data gap in the light curve simulation. As demonstrated in previous sections, MAGIC can recover the input parameters at high accuracy {for the majority of tested events}. In particular, \texttt{estimator} alone can achieve typical fractional errors of $\sim7\%$ and $\sim1.4\%$ for the binary mass ratio $q$ and the separation $s$, respectively. As demonstrated in the analysis of a real microlensing event (KB190371; \citealt{Kim:2021}), MAGIC is also capable of identifying degenerate models that explain the same data. The proposed method is fast to perform, with an estimated speedup of $\sim10^2$--$10^5$ depending on whether or not the optimization is needed.

While MAGIC is readily applicable to the analysis of real world data sets, a few improvements can be made to make MAGIC more powerful and more robust. 
First, this work focuses on binaries with relatively large mass ratios ($q>10^{-3}$), and extension into lower values may be necessary in order to apply MAGIC to the analysis of low-mass planetary events.
Second, the source scaled size $\rho$ is not inferred in the current version of MAGIC for reasons given in Section~\ref{sec:unet_data}, although it can still be constrained in the optimization stage (see Section~\ref{sec:application}). The inclusion of $\rho$ into MAGIC involves effectively both classification and regression tasks, as one will need to determine whether or not the finite source effect is detected and if yes, infer the value of $\rho$.
Furthermore, the inclusion of additional effects---especially the microlensing parallax effect \citep{Gould_parallax} and the binary source effect \citep{Griest:1992}---may be necessary in order for MAGIC to provide robust estimates on binaries with relatively weak signals.
Last but not least, MAGIC can also be applied to the analysis of microlensing events with more than two lenses, for which the sampling-based method becomes much more time-consuming \citep[e.g.,][]{Gaudi:2008, Kuang:2021}.

MAGIC uses neural CDE to handle irregular sampling and data gaps. Compared to other existing methods that have been used to handle the same issues in astronomical time series \citep[e.g.,][]{Charnock:2017,Naul:2018}, neural CDE takes directly the irregularly sampled time series and models the underlying process with a differential equation. 
{This approach has been demonstrated to achieve good performances in other disciplines that also involve time series data \citep[e.g.,][]{kidger2020neural}.}
As irregular sampling and data gaps are common in astronomical observations, we expect that neural CDE will be useful to many other fields in time domain astronomy.

%% file: app_large_chi2.tex
\begin{figure}
    \centering
    \includegraphics[width=\linewidth]{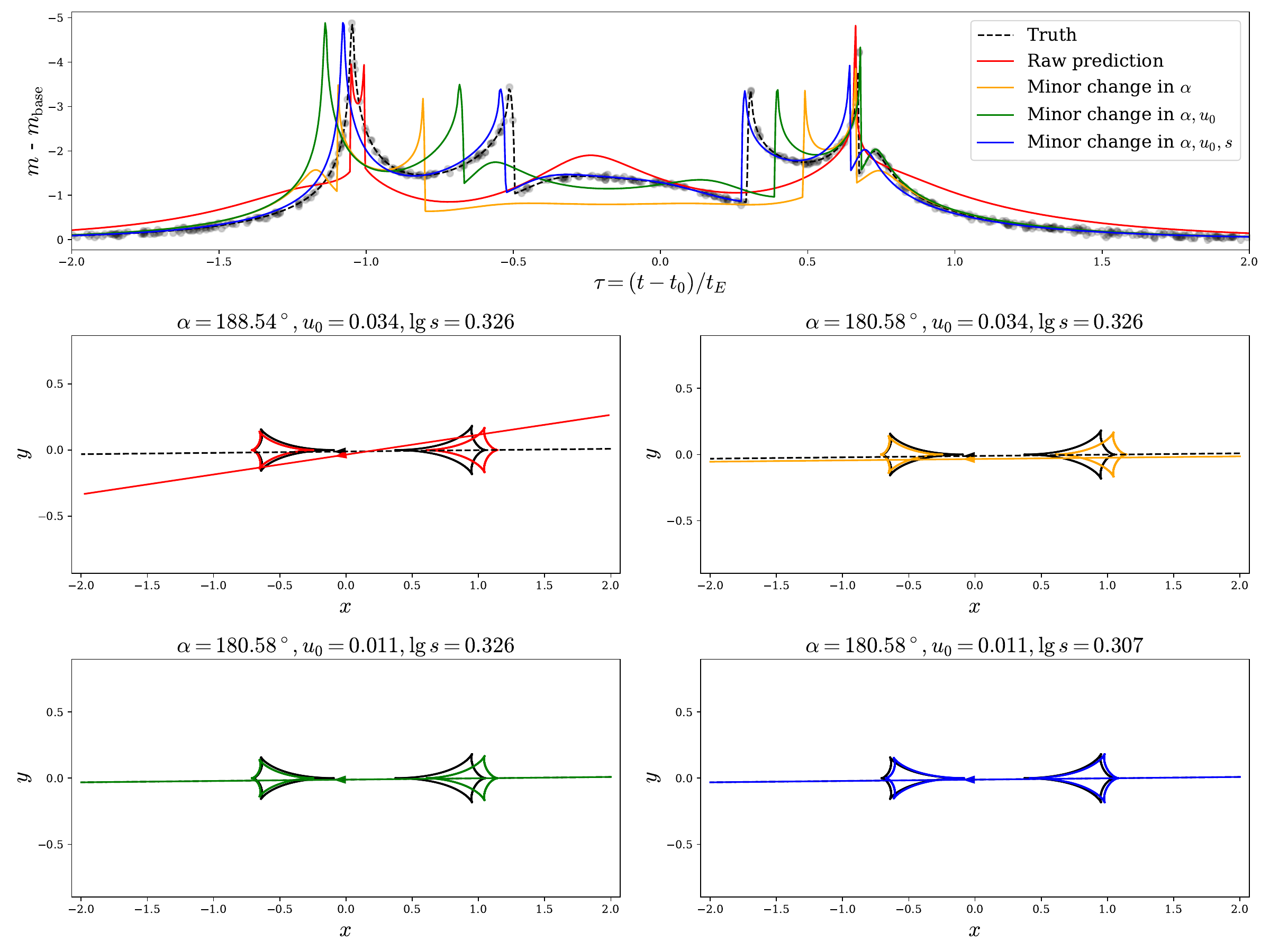}
    \caption{{An example event for which the raw prediction of \texttt{estimator} yields the largest $\chi^2$ value among all test events. The top panel shows the data and model light curves, and the middle and bottom panels are the corresponding microlensing geometry plots. The raw prediction of \texttt{estimator} is shown in red, and the ground truth is shown in black. We then change one by one the values of $\alpha$, $u_0$, and $s$ from the raw prediction to the ground truth values and show the corresponding light curves and lensing geometry plots with different colors.}}
    \label{fig:param_vs_lc}
\end{figure}

{As described in Section \ref{sec:ncde_perform}, \texttt{estimator} achieves reasonably high accuracy in microlensing parameters but fails to obtain good light curve fits (i.e. small $\chi^2$ values) in $\sim20\%-30\%$ of the test cases. This phenomenon arises because a small deviation in the parameter estimation may still lead to a large deviation in the light curve behavior, thus resulting in a large $\chi^2$ value. Here we examine a specific example, which is the event with the largest $\chi^2$ value in the test set. The predicted parameters only differ from the ground truth values by $(0.02, 0.04, 0.02, 0.04)$ in $(u_0, \lg q, \lg s, \alpha/180^\circ)$. However, the corresponding light curve deviates from the ground truth drastically, as shown by the red curve in the top panel of Figure~\ref{fig:param_vs_lc}. To illustrate the impact on the light curve behaviour caused by minor changes in the parameter values, we change one-by-one the values of $\alpha$, $u_0$, and $\lg s$ from the model prediction to the ground truth values, and plot the corresponding light curves in the same figure. 
We observe that the light curve fit remains bad until all the three parameters are fine tuned. This confirms that even when the parameter estimation is accurate enough, the light curve fit may still looks bad. Nevertheless, an accurate estimation of the parameters is enough for most applications.}

%% file: msV1.bbl
\begin{thebibliography}{}
\expandafter\ifx\csname natexlab\endcsname\relax\def\natexlab#1{#1}\fi
\providecommand{\url}[1]{\href{#1}{#1}}
\providecommand{\dodoi}[1]{doi:~\href{http://doi.org/#1}{\nolinkurl{#1}}}
\providecommand{\doeprint}[1]{\href{http://ascl.net/#1}{\nolinkurl{http://ascl.net/#1}}}
\providecommand{\doarXiv}[1]{\href{https://arxiv.org/abs/#1}{\nolinkurl{https://arxiv.org/abs/#1}}}

\bibitem[{{Albrow} {et~al.}(2000){Albrow}, {Beaulieu}, {Caldwell}, {DePoy},
  {Dominik}, {Gaudi}, {Gould}, {Greenhill}, {Hill}, {Kane}, {Martin},
  {Menzies}, {Naber}, {Pogge}, {Pollard}, {Sackett}, {Sahu}, {Vermaak},
  {Watson}, {Williams}, \& {PLANET Collaboration}}]{Albrow:2000}
{Albrow}, M.~D., {Beaulieu}, J.~P., {Caldwell}, J.~A.~R., {et~al.} 2000, \apj,
  535, 176, \dodoi{10.1086/308842}

\bibitem[{{An}(2005)}]{An:2005}
{An}, J.~H. 2005, \mnras, 356, 1409, \dodoi{10.1111/j.1365-2966.2004.08581.x}

\bibitem[{Bishop \& Nasrabadi(2006)}]{bishop2006pattern}
Bishop, C.~M., \& Nasrabadi, N.~M. 2006, Pattern recognition and machine
  learning (Springer)

\bibitem[{{Bozza}(2010)}]{Bozza:2010}
{Bozza}, V. 2010, \mnras, 408, 2188, \dodoi{10.1111/j.1365-2966.2010.17265.x}

\bibitem[{{Bozza} {et~al.}(2018){Bozza}, {Bachelet}, {Bartoli{\'c}}, {Heintz},
  {Hoag}, \& {Hundertmark}}]{Bozza:2018}
{Bozza}, V., {Bachelet}, E., {Bartoli{\'c}}, F., {et~al.} 2018, Monthly Notices
  of the Royal Astronomical Society, 479, 5157, \dodoi{10.1093/mnras/sty1791}

\bibitem[{{Charnock} \& {Moss}(2017)}]{Charnock:2017}
{Charnock}, T., \& {Moss}, A. 2017, Astrophysical Journal Letters, 837, L28,
  \dodoi{10.3847/2041-8213/aa603d}

\bibitem[{Che {et~al.}(2018)Che, Purushotham, Cho, Sontag, \&
  Liu}]{che2018recurrent}
Che, Z., Purushotham, S., Cho, K., Sontag, D., \& Liu, Y. 2018, Scientific
  reports, 8, 1

\bibitem[{{Chung} {et~al.}(2005){Chung}, {Han}, {Park}, {Kim}, {Kang}, {Ryu},
  {Kim}, {Jeon}, {Lee}, {Chang}, {Lee}, \& {Kang}}]{Chung:2005}
{Chung}, S.-J., {Han}, C., {Park}, B.-G., {et~al.} 2005, \apj, 630, 535,
  \dodoi{10.1086/432048}

\bibitem[{{Di Stefano} \& {Mao}(1996)}]{DiStefano:1996}
{Di Stefano}, R., \& {Mao}, S. 1996, \apj, 457, 93, \dodoi{10.1086/176713}

\bibitem[{{Dominik}(1999)}]{Dominik:1999}
{Dominik}, M. 1999, \aap, 349, 108.
\newblock \doarXiv{astro-ph/9903014}

\bibitem[{{Dong} {et~al.}(2006){Dong}, {DePoy}, {Gaudi}, {Gould}, {Han},
  {Park}, {Pogge}, {MuFun Collaboration}, {Udalski}, {Szewczyk}, {Kubiak},
  {Szyma{\'n}ski}, {Pietrzy{\'n}ski}, {Soszy{\'n}ski}, {Wyrzykowski},
  {{\.Z}ebru{\'n}}, \& {OGLE Collaboration}}]{Dong:2006}
{Dong}, S., {DePoy}, D.~L., {Gaudi}, B.~S., {et~al.} 2006, \apj, 642, 842,
  \dodoi{10.1086/501224}

\bibitem[{{Einstein}(1936)}]{Einstein:1936}
{Einstein}, A. 1936, Science, 84, 506, \dodoi{10.1126/science.84.2188.506}

\bibitem[{Foreman-Mackey(2016)}]{corner}
Foreman-Mackey, D. 2016, The Journal of Open Source Software, 1, 24,
  \dodoi{10.21105/joss.00024}

\bibitem[{Gaudi(2012)}]{gaudi2012microlensing}
Gaudi, B.~S. 2012, Annual Review of Astronomy and Astrophysics, 50, 411

\bibitem[{{Gaudi} {et~al.}(2008){Gaudi}, {Bennett}, {Udalski}, {Gould},
  {Christie}, {Maoz}, {Dong}, {McCormick}, {Szyma{\'n}ski}, {Tristram},
  {Nikolaev}, {Paczy{\'n}ski}, {Kubiak}, {Pietrzy{\'n}ski}, {Soszy{\'n}ski},
  {Szewczyk}, {Ulaczyk}, {Wyrzykowski}, {OGLE Collaboration}, {DePoy}, {Han},
  {Kaspi}, {Lee}, {Mallia}, {Natusch}, {Pogge}, {Park}, {{\ensuremath{\mu}}-Fun
  Collabortion}, {Abe}, {Bond}, {Botzler}, {Fukui}, {Hearnshaw}, {Itow},
  {Kamiya}, {Korpela}, {Kilmartin}, {Lin}, {Masuda}, {Matsubara}, {Motomura},
  {Muraki}, {Nakamura}, {Okumura}, {Ohnishi}, {Rattenbury}, {Sako}, {Saito},
  {Sato}, {Skuljan}, {Sullivan}, {Sumi}, {Sweatman}, {Yock}, {MOA
  Collaboration}, {Albrow}, {Allan}, {Beaulieu}, {Burgdorf}, {Cook},
  {Coutures}, {Dominik}, {Dieters}, {Fouqu{\'e}}, {Greenhill}, {Horne},
  {Steele}, {Tsapras}, {Planet Collaboration}, {RoboNet Collaborations},
  {Chaboyer}, {Crocker}, {Frank}, \& {Macintosh}}]{Gaudi:2008}
{Gaudi}, B.~S., {Bennett}, D.~P., {Udalski}, A., {et~al.} 2008, Science, 319,
  927, \dodoi{10.1126/science.1151947}

\bibitem[{Godines {et~al.}(2019)Godines, Bachelet, Narayan, \&
  Street}]{GODINES2019100298}
Godines, D., Bachelet, E., Narayan, G., \& Street, R. 2019, Astronomy and
  Computing, 28, 100298, \dodoi{https://doi.org/10.1016/j.ascom.2019.100298}

\bibitem[{Goodfellow {et~al.}(2016)Goodfellow, Bengio, \&
  Courville}]{goodfellow2016deep}
Goodfellow, I., Bengio, Y., \& Courville, A. 2016, Deep learning (MIT press)

\bibitem[{{Gould}(1992)}]{Gould_parallax}
{Gould}, A. 1992, \apj, 392, 442, \dodoi{10.1086/171443}

\bibitem[{{Gould}(1995)}]{Gould:1995}
---. 1995, \apj, 440, 510, \dodoi{10.1086/175292}

\bibitem[{{Gould} {et~al.}(2010{\natexlab{a}}){Gould}, {Dong}, {Bennett},
  {Bond}, {Udalski}, \& {Kozlowski}}]{Gould:2010a}
{Gould}, A., {Dong}, S., {Bennett}, D.~P., {et~al.} 2010{\natexlab{a}}, \apj,
  710, 1800, \dodoi{10.1088/0004-637X/710/2/1800}

\bibitem[{{Gould} \& {Han}(2000)}]{GouldHan:2000}
{Gould}, A., \& {Han}, C. 2000, \apj, 538, 653, \dodoi{10.1086/309180}

\bibitem[{{Gould} \& {Loeb}(1992)}]{Gould:1992}
{Gould}, A., \& {Loeb}, A. 1992, \apj, 396, 104, \dodoi{10.1086/171700}

\bibitem[{{Gould} {et~al.}(2010{\natexlab{b}}){Gould}, {Dong}, {Gaudi},
  {Udalski}, {Bond}, {Greenhill}, {Street}, {Dominik}, {Sumi}, {Szyma{\'n}ski},
  {Han}, {Allen}, {Bolt}, {Bos}, {Christie}, {DePoy}, {Drummond}, {Eastman},
  {Gal-Yam}, {Higgins}, {Janczak}, {Kaspi}, {Koz{\l}owski}, {Lee}, {Mallia},
  {Maury}, {Maoz}, {McCormick}, {Monard}, {Moorhouse}, {Morgan}, {Natusch},
  {Ofek}, {Park}, {Pogge}, {Polishook}, {Santallo}, {Shporer}, {Spector},
  {Thornley}, {Yee}, {{\ensuremath{\mu}}FUN Collaboration}, {Kubiak},
  {Pietrzy{\'n}ski}, {Soszy{\'n}ski}, {Szewczyk}, {Wyrzykowski}, {Ulaczyk},
  {Poleski}, {OGLE Collaboration}, {Abe}, {Bennett}, {Botzler}, {Douchin},
  {Freeman}, {Fukui}, {Furusawa}, {Hearnshaw}, {Hosaka}, {Itow}, {Kamiya},
  {Kilmartin}, {Korpela}, {Lin}, {Ling}, {Makita}, {Masuda}, {Matsubara},
  {Miyake}, {Muraki}, {Nagaya}, {Nishimoto}, {Ohnishi}, {Okumura}, {Perrott},
  {Philpott}, {Rattenbury}, {Saito}, {Sako}, {Sullivan}, {Sweatman},
  {Tristram}, {von Seggern}, {Yock}, {MOA Collaboration}, {Albrow}, {Batista},
  {Beaulieu}, {Brillant}, {Caldwell}, {Calitz}, {Cassan}, {Cole}, {Cook},
  {Coutures}, {Dieters}, {Dominis Prester}, {Donatowicz}, {Fouqu{\'e}}, {Hill},
  {Hoffman}, {Jablonski}, {Kane}, {Kains}, {Kubas}, {Marquette}, {Martin},
  {Martioli}, {Meintjes}, {Menzies}, {Pedretti}, {Pollard}, {Sahu}, {Vinter},
  {Wambsganss}, {Watson}, {Williams}, {Zub}, {PLANET Collaboration}, {Allan},
  {Bode}, {Bramich}, {Burgdorf}, {Clay}, {Fraser}, {Hawkins}, {Horne},
  {Kerins}, {Lister}, {Mottram}, {Saunders}, {Snodgrass}, {Steele}, {Tsapras},
  {RoboNet Collaboration}, {J{\o}rgensen}, {Anguita}, {Bozza}, {Calchi Novati},
  {Harps{\o}e}, {Hinse}, {Hundertmark}, {Kj{\ae}rgaard}, {Liebig}, {Mancini},
  {Masi}, {Mathiasen}, {Rahvar}, {Ricci}, {Scarpetta}, {Southworth}, {Surdej},
  {Th{\"o}ne}, \& {MiNDSTEp Consortium}}]{Gould:2010}
{Gould}, A., {Dong}, S., {Gaudi}, B.~S., {et~al.} 2010{\natexlab{b}}, \apj,
  720, 1073, \dodoi{10.1088/0004-637X/720/2/1073}

\bibitem[{{Griest} \& {Hu}(1992)}]{Griest:1992}
{Griest}, K., \& {Hu}, W. 1992, \apj, 397, 362, \dodoi{10.1086/171793}

\bibitem[{{Griest} \& {Safizadeh}(1998)}]{Griest:1998}
{Griest}, K., \& {Safizadeh}, N. 1998, \apj, 500, 37, \dodoi{10.1086/305729}

\bibitem[{Harris {et~al.}(2020)Harris, Millman, van~der Walt, Gommers,
  Virtanen, Cournapeau, Wieser, Taylor, Berg, Smith, Kern, Picus, Hoyer, van
  Kerkwijk, Brett, Haldane, del R{\'{i}}o, Wiebe, Peterson,
  G{\'{e}}rard-Marchant, Sheppard, Reddy, Weckesser, Abbasi, Gohlke, \&
  Oliphant}]{harris2020array}
Harris, C.~R., Millman, K.~J., van~der Walt, S.~J., {et~al.} 2020, Nature, 585,
  357, \dodoi{10.1038/s41586-020-2649-2}

\bibitem[{He {et~al.}(2015)He, Zhang, Ren, \& Sun}]{he2015prelu}
He, K., Zhang, X., Ren, S., \& Sun, J. 2015, arXiv preprint arXiv:1502.01852

\bibitem[{He {et~al.}(2016)He, Zhang, Ren, \& Sun}]{he2016deep}
He, K., Zhang, X., Ren, S., \& Sun, J. 2016, in Proceedings of the IEEE
  conference on computer vision and pattern recognition, 770--778

\bibitem[{Hunter(2007)}]{Hunter2007matplotlib}
Hunter, J.~D. 2007, Computing in Science \& Engineering, 9, 90,
  \dodoi{10.1109/MCSE.2007.55}

\bibitem[{{Hwang} {et~al.}(2018){Hwang}, {Udalski}, {Shvartzvald}, {Ryu},
  {Albrow}, {Chung}, {Gould}, {Han}, {Jung}, {Shin}, {Yee}, {Zhu}, {Cha},
  {Kim}, {Kim}, {Kim}, {Lee}, {Lee}, {Lee}, {Park}, {Pogge}, {KMTNet
  Collaboration}, {Skowron}, {Mr{\'o}z}, {Poleski}, {Koz{\l}owski},
  {Soszy{\'n}ski}, {Pietrukowicz}, {Szyma{\'n}ski}, {Ulaczyk}, {Pawlak}, {OGLE
  Collaboration}, {Bryden}, {Beichman}, {Calchi Novati}, {Gaudi}, {Henderson},
  {Jacklin}, {Penny}, \& {UKIRT Microlensing Team}}]{Hwang:2018}
{Hwang}, K.~H., {Udalski}, A., {Shvartzvald}, Y., {et~al.} 2018, \aj, 155, 20,
  \dodoi{10.3847/1538-3881/aa992f}

\bibitem[{{Kennedy} {et~al.}(2021){Kennedy}, {Nash}, {Rattenbury}, \&
  {Kempa-Liehr}}]{Kennedy:2021}
{Kennedy}, A., {Nash}, G., {Rattenbury}, N.~J., \& {Kempa-Liehr}, A.~W. 2021,
  Astronomy and Computing, 35, 100460, \dodoi{10.1016/j.ascom.2021.100460}

\bibitem[{{Khakpash} {et~al.}(2019){Khakpash}, {Penny}, \&
  {Pepper}}]{Khakpash:2019}
{Khakpash}, S., {Penny}, M., \& {Pepper}, J. 2019, \aj, 158, 9,
  \dodoi{10.3847/1538-3881/ab1fe3}

\bibitem[{Kidger(2022)}]{kidger_thesis}
Kidger, P. 2022, arXiv preprint arXiv:2202.02435

\bibitem[{Kidger {et~al.}(2020)Kidger, Morrill, Foster, \&
  Lyons}]{kidger2020neural}
Kidger, P., Morrill, J., Foster, J., \& Lyons, T. 2020, Advances in Neural
  Information Processing Systems, 33, 6696

\bibitem[{{Kim} {et~al.}(2016){Kim}, {Lee}, {Park}, {Kim}, {Cha}, {Lee}, {Han},
  {Chun}, \& {Yuk}}]{Kim:2016}
{Kim}, S.-L., {Lee}, C.-U., {Park}, B.-G., {et~al.} 2016, Journal of Korean
  Astronomical Society, 49, 37, \dodoi{10.5303/JKAS.2016.49.1.037}

\bibitem[{{Kim} {et~al.}(2021){Kim}, {Chung}, {Yee}, {Udalski}, {Bond}, {Jung},
  {Gould}, {Albrow}, {Han}, {Hwang}, {Ryu}, {Shin}, {Shvartzvald}, {Zang},
  {Cha}, {Kim}, {Kim}, {Kim}, {Lee}, {Lee}, {Lee}, {Park}, {Pogge}, {KMTNet
  Collaboration}, {Poleski}, {Mr{\'o}z}, {Skowron}, {Szyma{\'n}ski},
  {Soszy{\'n}ski}, {Pietrukowicz}, {Koz{\l}owski}, {Ulaczyk}, {Rybicki},
  {Iwanek}, {Wrona}, {Gromadzki}, {OGLE Collaboration}, {Abe}, {Barry},
  {Bennett}, {Bhattacharya}, {Donachie}, {Fujii}, {Fukui}, {Itow}, {Hirao},
  {Kirikawa}, {Kondo}, {Koshimoto}, {Matsubara}, {Muraki}, {Miyazaki}, {Ranc},
  {Rattenbury}, {Satoh}, {Shoji}, {Sumi}, {Suzuki}, {Tristram}, {Tanaka},
  {Yamawaki}, {Yonehara}, \& {MOA Collaboration}}]{Kim:2021}
{Kim}, Y.~H., {Chung}, S.-J., {Yee}, J.~C., {et~al.} 2021, \aj, 162, 17,
  \dodoi{10.3847/1538-3881/abf930}

\bibitem[{Kingma \& Ba(2014)}]{kingma2014adam}
Kingma, D.~P., \& Ba, J. 2014, arXiv preprint arXiv:1412.6980

\bibitem[{Kluyver {et~al.}(2016)Kluyver, Ragan-Kelley, P{\'e}rez, Granger,
  Bussonnier, Frederic, Kelley, Hamrick, Grout, Corlay, Ivanov, Avila, Abdalla,
  Willing, \& development team}]{jupyter}
Kluyver, T., Ragan-Kelley, B., P{\'e}rez, F., {et~al.} 2016, in Positioning and
  Power in Academic Publishing: Players, Agents and Agendas, ed. F.~Loizides \&
  B.~Scmidt (IOS Press), 87--90.
\newblock \url{https://eprints.soton.ac.uk/403913/}

\bibitem[{{Kuang} {et~al.}(2021){Kuang}, {Mao}, {Wang}, {Zang}, \&
  {Long}}]{Kuang:2021}
{Kuang}, R., {Mao}, S., {Wang}, T., {Zang}, W., \& {Long}, R.~J. 2021, \mnras,
  503, 6143, \dodoi{10.1093/mnras/stab509}

\bibitem[{{Makinen} {et~al.}(2021){Makinen}, {Lancaster},
  {Villaescusa-Navarro}, {Melchior}, {Ho}, {Perreault-Levasseur}, \&
  {Spergel}}]{Makinen:2021}
{Makinen}, T.~L., {Lancaster}, L., {Villaescusa-Navarro}, F., {et~al.} 2021,
  \jcap, 2021, 081, \dodoi{10.1088/1475-7516/2021/04/081}

\bibitem[{{Mao} \& {Paczynski}(1991)}]{Mao:1991}
{Mao}, S., \& {Paczynski}, B. 1991, Astrophysical Journal Letters, 374, L37,
  \dodoi{10.1086/186066}

\bibitem[{Morrill {et~al.}(2021)Morrill, Salvi, Kidger, \&
  Foster}]{morrill2021neural}
Morrill, J., Salvi, C., Kidger, P., \& Foster, J. 2021, in International
  Conference on Machine Learning, PMLR, 7829--7838

\bibitem[{{Mr{\'o}z}(2020)}]{mroz2020ident}
{Mr{\'o}z}, P. 2020, Acta Astronomica, 70, 169,
  \dodoi{10.32023/0001-5237/70.3.1}

\bibitem[{{Mr{\'o}z} {et~al.}(2017){Mr{\'o}z}, {Udalski}, {Skowron}, {Poleski},
  {Koz{\l}owski}, {Szyma{\'n}ski}, {Soszy{\'n}ski}, {Wyrzykowski},
  {Pietrukowicz}, {Ulaczyk}, {Skowron}, \& {Pawlak}}]{Mroz:2017}
{Mr{\'o}z}, P., {Udalski}, A., {Skowron}, J., {et~al.} 2017, \nat, 548, 183,
  \dodoi{10.1038/nature23276}

\bibitem[{{Naul} {et~al.}(2018){Naul}, {Bloom}, {P{\'e}rez}, \& {van der
  Walt}}]{Naul:2018}
{Naul}, B., {Bloom}, J.~S., {P{\'e}rez}, F., \& {van der Walt}, S. 2018, Nature
  Astronomy, 2, 151, \dodoi{10.1038/s41550-017-0321-z}

\bibitem[{Nelder \& Mead(1965)}]{Nelder:1965}
Nelder, J.~A., \& Mead, R. 1965, Comput. J., 7, 308

\bibitem[{{Paczynski}(1986)}]{Paczynski:1986}
{Paczynski}, B. 1986, The Astrophysical Journal, 304, 1, \dodoi{10.1086/164140}

\bibitem[{Paszke {et~al.}(2019)Paszke, Gross, Massa, Lerer, Bradbury, Chanan,
  Killeen, Lin, Gimelshein, Antiga, Desmaison, Kopf, Yang, DeVito, Raison,
  Tejani, Chilamkurthy, Steiner, Fang, Bai, \& Chintala}]{pytorch}
Paszke, A., Gross, S., Massa, F., {et~al.} 2019, in Advances in Neural
  Information Processing Systems 32 (Curran Associates, Inc.), 8024--8035

\bibitem[{Penny(2014)}]{penny2014speeding}
Penny, M.~T. 2014, The Astrophysical Journal, 790, 142

\bibitem[{{Penny} {et~al.}(2019){Penny}, {Gaudi}, {Kerins}, {Rattenbury},
  {Mao}, {Robin}, \& {Calchi Novati}}]{Penny:2019}
{Penny}, M.~T., {Gaudi}, B.~S., {Kerins}, E., {et~al.} 2019, The Astrophysical
  Journal Supplement Series, 241, 3, \dodoi{10.3847/1538-4365/aafb69}

\bibitem[{{Poleski} \& {Yee}(2019)}]{mulensmodel}
{Poleski}, R., \& {Yee}, J.~C. 2019, Astronomy and Computing, 26, 35,
  \dodoi{10.1016/j.ascom.2018.11.001}

\bibitem[{Ronneberger {et~al.}(2015)Ronneberger, Fischer, \&
  Brox}]{ronneberger2015unet}
Ronneberger, O., Fischer, P., \& Brox, T. 2015, in International Conference on
  Medical image computing and computer-assisted intervention, Springer,
  234--241

\bibitem[{Rubanova {et~al.}(2019)Rubanova, Chen, \&
  Duvenaud}]{rubanova2019latent}
Rubanova, Y., Chen, R.~T., \& Duvenaud, D.~K. 2019, Advances in neural
  information processing systems, 32

\bibitem[{Siddique {et~al.}(2021)Siddique, Paheding, Elkin, \&
  Devabhaktuni}]{Siddique2021unet_review}
Siddique, N., Paheding, S., Elkin, C.~P., \& Devabhaktuni, V. 2021, IEEE
  Access, 9, 82031, \dodoi{10.1109/ACCESS.2021.3086020}

\bibitem[{{Skowron} {et~al.}(2011){Skowron}, {Udalski}, {Gould}, {Dong},
  {Monard}, {Han}, {Nelson}, {McCormick}, {Moorhouse}, {Thornley}, {Maury},
  {Bramich}, {Greenhill}, {Koz{\l}owski}, {Bond}, {Poleski}, {Wyrzykowski},
  {Ulaczyk}, {Kubiak}, {Szyma{\'n}ski}, {Pietrzy{\'n}ski}, {Soszy{\'n}ski},
  {OGLE Collaboration}, {Gaudi}, {Yee}, {Hung}, {Pogge}, {DePoy}, {Lee},
  {Park}, {Allen}, {Mallia}, {Drummond}, {Bolt}, {{\ensuremath{\mu}}FUN
  Collaboration}, {Allan}, {Browne}, {Clay}, {Dominik}, {Fraser}, {Horne},
  {Kains}, {Mottram}, {Snodgrass}, {Steele}, {Street}, {Tsapras}, {RoboNet
  Collaboration}, {Abe}, {Bennett}, {Botzler}, {Douchin}, {Freeman}, {Fukui},
  {Furusawa}, {Hayashi}, {Hearnshaw}, {Hosaka}, {Itow}, {Kamiya}, {Kilmartin},
  {Korpela}, {Lin}, {Ling}, {Makita}, {Masuda}, {Matsubara}, {Muraki},
  {Nagayama}, {Miyake}, {Nishimoto}, {Ohnishi}, {Perrott}, {Rattenbury},
  {Saito}, {Skuljan}, {Sullivan}, {Sumi}, {Suzuki}, {Sweatman}, {Tristram},
  {Wada}, {Yock}, {MOA Collaboration}, {Beaulieu}, {Fouqu{\'e}}, {Albrow},
  {Batista}, {Brillant}, {Caldwell}, {Cassan}, {Cole}, {Cook}, {Coutures},
  {Dieters}, {Dominis Prester}, {Donatowicz}, {Kane}, {Kubas}, {Marquette},
  {Martin}, {Menzies}, {Sahu}, {Wambsganss}, {Williams}, {Zub}, \& {PLANET
  Collaboration}}]{Skowron:2011}
{Skowron}, J., {Udalski}, A., {Gould}, A., {et~al.} 2011, \apj, 738, 87,
  \dodoi{10.1088/0004-637X/738/1/87}

\bibitem[{{Song} {et~al.}(2014){Song}, {Mao}, \& {An}}]{Song:2014}
{Song}, Y.-Y., {Mao}, S., \& {An}, J.~H. 2014, \mnras, 437, 4006,
  \dodoi{10.1093/mnras/stt2222}

\bibitem[{Sudre {et~al.}(2017)Sudre, Li, Vercauteren, Ourselin, \&
  Jorge~Cardoso}]{sudre2017dice}
Sudre, C.~H., Li, W., Vercauteren, T., Ourselin, S., \& Jorge~Cardoso, M. 2017,
  in Deep learning in medical image analysis and multimodal learning for
  clinical decision support (Springer), 240--248

\bibitem[{{Van Oort} {et~al.}(2019){Van Oort}, {Xu}, {Offner}, \&
  {Gutermuth}}]{VanOort:2019}
{Van Oort}, C.~M., {Xu}, D., {Offner}, S. S.~R., \& {Gutermuth}, R.~A. 2019,
  \apj, 880, 83, \dodoi{10.3847/1538-4357/ab275e}

\bibitem[{Vermaak(2003)}]{verm2003rapid}
Vermaak, P. 2003, Monthly Notices of the Royal Astronomical Society, 344, 651,
  \dodoi{10.1046/j.1365-8711.2003.06851.x}

\bibitem[{Virtanen {et~al.}(2020)Virtanen, Gommers, Oliphant, Haberland, Reddy,
  Cournapeau, Burovski, Peterson, Weckesser, Bright, {van der Walt}, Brett,
  Wilson, Millman, Mayorov, Nelson, Jones, Kern, Larson, Carey, Polat, Feng,
  Moore, {VanderPlas}, Laxalde, Perktold, Cimrman, Henriksen, Quintero, Harris,
  Archibald, Ribeiro, Pedregosa, {van Mulbregt}, \& {SciPy 1.0
  Contributors}}]{2020SciPy-NMeth}
Virtanen, P., Gommers, R., Oliphant, T.~E., {et~al.} 2020, Nature Methods, 17,
  261, \dodoi{10.1038/s41592-019-0686-2}

\bibitem[{{Wang} {et~al.}(2022){Wang}, {Zang}, {Zhu}, {Hwang}, {Udalski},
  {Gould}, {Han}, {Albrow}, {Chung}, {Jung}, {Kim}, {Ryu}, {Shin},
  {Shvartzvald}, {Yee}, {Cha}, {Kim}, {Kim}, {Kim}, {Lee}, {Lee}, {Lee},
  {Park}, {Pogge}, {Poleski}, {Mr{\'o}z}, {Skowron}, {Szyma{\'n}ski},
  {Soszy{\'n}ski}, {Pietrukowicz}, {Koz{\l}owski}, {Ulaczyk}, {Rybicki},
  {Iwanek}, {Wrona}, {Gromadzki}, {Yang}, {Mao}, \& {Zhang}}]{Wang:2022}
{Wang}, H., {Zang}, W., {Zhu}, W., {et~al.} 2022, \mnras, 510, 1778,
  \dodoi{10.1093/mnras/stab3581}

\bibitem[{{W}es {M}c{K}inney(2010)}]{mckinney-proc-scipy-2010-pandas}
{W}es {M}c{K}inney. 2010, in {P}roceedings of the 9th {P}ython in {S}cience
  {C}onference, ed. {S}t\'efan van~der {W}alt \& {J}arrod {M}illman, 56 -- 61,
  \dodoi{10.25080/Majora-92bf1922-00a}

\bibitem[{Wyrzykowski {et~al.}(2015)Wyrzykowski, Rynkiewicz, Skowron,
  Koz{\l}owski, Udalski, Szyma{\'n}ski, Kubiak, Soszy{\'n}ski, Pietrzy{\'n}ski,
  Poleski, {et~al.}}]{wyrzykowski2015ogle}
Wyrzykowski, {\L}., Rynkiewicz, A.~E., Skowron, J., {et~al.} 2015, The
  Astrophysical Journal Supplement Series, 216, 12

\bibitem[{{Yoo} {et~al.}(2004){Yoo}, {DePoy}, {Gal-Yam}, {Gaudi}, {Gould},
  {Han}, {Lipkin}, {Maoz}, {Ofek}, {Park}, {Pogge}, {Mu-Fun Collaboration},
  {Udalski}, {Soszy{\'n}ski}, {Wyrzykowski}, {Kubiak}, {Szyma{\'n}ski},
  {Pietrzy{\'n}ski}, {Szewczyk}, {{\.Z}ebru{\'n}}, \& {OGLE
  Collaboration}}]{Yoo:2004}
{Yoo}, J., {DePoy}, D.~L., {Gal-Yam}, A., {et~al.} 2004, \apj, 603, 139,
  \dodoi{10.1086/381241}

\bibitem[{{Zang} {et~al.}(2022){Zang}, {Yang}, {Han}, {Lee}, {Udalski},
  {Gould}, {Mao}, {Zhang}, {Zhu}, {Albrow}, {Chung}, {Hwang}, {Jung}, {Ryu},
  {Shin}, {Shvartzvald}, {Yee}, {Cha}, {Kim}, {Kim}, {Kim}, {Lee}, {Lee},
  {Park}, {Mr{\'o}z}, {Skowron}, {Poleski}, {Szyma{\'n}ski}, {Soszy{\'n}ski},
  {Pietrukowicz}, {Koz{\l}owski}, {Ulaczyk}, {Rybicki}, {Iwanek}, {Wrona}, \&
  {Gromadzki}}]{Zang:2022}
{Zang}, W., {Yang}, H., {Han}, C., {et~al.} 2022, arXiv e-prints,
  arXiv:2204.02017.
\newblock \doarXiv{2204.02017}

\bibitem[{Zhang {et~al.}(2021)Zhang, Bloom, Gaudi, Lanusse, Lam, \&
  Lu}]{zhang2021real}
Zhang, K., Bloom, J.~S., Gaudi, B.~S., {et~al.} 2021, The Astronomical Journal,
  161, 262

\bibitem[{{Zhang} {et~al.}(2022){Zhang}, {Gaudi}, \& {Bloom}}]{Zhang:2022}
{Zhang}, K., {Gaudi}, B.~S., \& {Bloom}, J.~S. 2022, Nature Astronomy, 6, 782,
  \dodoi{10.1038/s41550-022-01671-6}

\bibitem[{{Zhu} \& {Dong}(2021)}]{Zhu:2021}
{Zhu}, W., \& {Dong}, S. 2021, \araa, 59,
  \dodoi{10.1146/annurev-astro-112420-020055}

\bibitem[{{Zhu} {et~al.}(2014){Zhu}, {Penny}, {Mao}, {Gould}, \&
  {Gendron}}]{Zhu:2014}
{Zhu}, W., {Penny}, M., {Mao}, S., {Gould}, A., \& {Gendron}, R. 2014, \apj,
  788, 73, \dodoi{10.1088/0004-637X/788/1/73}

\bibitem[{{Zhu} {et~al.}(2017){Zhu}, {Udalski}, {Novati}, {Chung}, {Jung},
  {Ryu}, {Shin}, {Gould}, {Lee}, {Albrow}, {Yee}, {Han}, {Hwang}, {Cha}, {Kim},
  {Kim}, {Kim}, {Kim}, {Lee}, {Park}, {Pogge}, {KMTNet Collaboration},
  {Poleski}, {Mr{\'o}z}, {Pietrukowicz}, {Skowron}, {Szyma{\'n}ski},
  {KozLowski}, {Ulaczyk}, {Pawlak}, {OGLE Collaboration}, {Beichman}, {Bryden},
  {Carey}, {Fausnaugh}, {Gaudi}, {Henderson}, {Shvartzvald}, {Wibking}, \&
  {Spitzer Team}}]{Zhu:2017}
{Zhu}, W., {Udalski}, A., {Novati}, S.~C., {et~al.} 2017, \aj, 154, 210,
  \dodoi{10.3847/1538-3881/aa8ef1}

\end{thebibliography}
